\documentclass{article}
\usepackage{graphicx} % Required for inserting images

\usepackage{amssymb,textcase,setspace,fancyhdr,enumerate,amsmath,physics,bm,color,enumitem}
\usepackage[sorting=none,backend=bibtex]{biblatex}
\usepackage{tikz}
\usetikzlibrary{arrows}
\usetikzlibrary{patterns,arrows,decorations.pathreplacing}
\usepackage{xparse}
\usepackage[left=1.5cm, right=2cm]{geometry}

\usepackage{xcolor}
\usepackage[urlcolor=blue]{hyperref}      % links to citations
\hypersetup{
    colorlinks = true,                    % text and not border
    citecolor = {blue},
    linkcolor = {purple},
           }

\newcommand{\vect}[1]{\mathbf{#1}}
\newcommand{\fixp}[2][]{%
  \vect{#2}^{\ast}%
  \if\relax\detokenize{#1}\relax
  \else(#1)\fi
}
\NewDocumentCommand{\fixpval}{m o o}{%
  #1^{\ast}%
  \IfValueT{#3}{_{#3}}%
  \IfValueT{#2}{(#2)}%
}

\title{Maximally chaotic competition for attention
in the cultural domain}
\author{
András Rusu$^{1,2}$ \and
Claudius Gros$^{3}$ \and
Bulcsú Sándor$^{1,2}$
\\[0.5em]
\small $^{1}$Department of Physics, Babe\c{s}-Bolyai University, Cluj-Napoca, Romania 
\\
\small $^{2}$Transylvanian Institute of Neuroscience, Cluj-Napoca, Romania
\\
\small $^{3}$Institute for Theoretical Physics, Goethe University Frankfurt, Frankfurt am Main, Germany
}

\date{}
\begin{document}

\maketitle

\begin{abstract}	
Memory is a key determinant when cultural items 
compete for attention and, consequently, 
for success, as in the case of songs on a music chart.
For modeling, one adds memory to 
Lotka-Volterra models, the reference for 
Markovian competitive processes. Here we
treat memory in terms of an exponential 
moving average, finding that it leads
to an extended region of winnerless chaos characterized by log-normal popularity 
statistics in trailing top-$k$ charts. 
Importantly, the observed log-normal behavior 
collapses to a power-law distribution when the 
feedback dynamics is fast on the scale of the 
charting period. This result is in agreement 
with the observed statistics of real-world music 
charts (e.g., Billboard and Spotify).

In a chaotic state, the size of the largest Lyapunov
exponent is a measure of how unpredictable the system
is. We find that the largest Lyapunov exponent varies 
strongly as a function of parameters in the phase where
winnerless chaos is stable. Interestingly, the
sets of parameters obtained by comparing simulations 
with real-world cultural-item dynamics extracted 
from Google Books and Google Trends, movies, Reddit, 
Wikipedia, Twitter, and scientific publications,
are located close to the points
in parameter space where the largest Lyapunov 
exponent reaches its local maximum,
namely, close to the point of maximal unpredictability.
This result suggests that cultural competitive 
processes are maximally chaotic when memory is
a key determinant.

\end{abstract}
%%%%%%%%%%%%%%%%%%%%%%%%%%%%%%%%%%%%%%%%%%%%%%%%%%%%%

%%%%%%%%%%%%%%%%%%%%%%
\section{Introduction}
%%%%%%%%%%%%%%%%%%%%%%

The dynamics of public discussion surrounding 
popular cultural items is typically characterized 
by unpredictable variations across several orders 
of magnitude: our attention is constantly redirected 
from one topic (e.g., news events or products of 
popular culture) to another, resulting in rapidly 
alternating patterns with sharp peaks when viewed 
as a time series~\cite{attention_dynamics}. This activity 
can be measured online through several proxies, such 
as the number of posts, likes, or shares on social 
media, as well as headlines in the press, allowing 
for the study of the statistical properties of
engagement.

It has been shown that, in certain popularity rankings 
of cultural items, the lifetime distribution (e.g., 
of music albums) has shifted from log-normal scaling 
to a power law over the last 40
years~\cite{Schneider2019,Schneider2021}. 
A possible cause is the acceleration of 
popularity dynamics: the timescale on which 
users interact with content 
has changed dramatically since the advent of
the internet.

%----------------------------------------
\subsection{Modeling attention dynamics}
%----------------------------------------

Large interconnected systems are often modeled using
generalized LV equations. Here, multiple dynamical 
phases can arise depending on the coupling
structure and the strength of the interactions. 
This is the case for Eq.~(\ref{eq:rescaled}); however, other routes to complexity are also possible.

For example, complexity may be induced
in large LV systems by asymmetric and/or disordered
interactions~\cite{Bunin1,Bunin2,GenerLV}, 
by the external environmental
noise~\cite{Opper92}, or by 
migration~\cite{Bunin1,Bunin3}.  Chaotic
attractors in asymmetric systems \cite{Vano4D},
as well as switching dynamics induced by 
instabilities — where trajectories temporarily 
visit saddle points connected by
heteroclinic connections~\cite{Ashwin2005,Voit2019}
-- have been reported in various contexts, 
ranging from motor activities driven by
chaotic neural networks~\cite{Varona2002} to animal search
patterns~\cite{Gutirrez2015}.

Other studies aimed to determine analytically the 
conditions under which winnerless chaos (WLC) occurs, 
such as in discrete LV variants, often assuming
some form of special symmetry in the interaction
matrix~\cite{GonzlezDaz2013}.

To understand how and why the dynamics of
cultural attention has changed over time, a
suitable model is required. In this regard, 
the attention economy of online social 
networks~\cite{heitmayer2025second} and the 
dynamics of information spreading have 
been studied using various
approaches, including probabilistic
models~\cite{OSN_stoch1,stoch2,stoch3}, dynamical processes on
networks~\cite{Bipartite,network2}, and differential equation
models~\cite{radicalization,word_usage,spanish_election}.

In particular, the acceleration of collective 
attention allocation has been demonstrated 
within the framework of a competitive
Lotka-Volterra (LV) model with memory, in 
which items compete for a limited amount of 
attention capacity, similarly to population
dynamics in biological systems. 
Within this approach to attention dynamics, 
acceleration corresponds to increased production 
and consumption rates of popular topics and
cultural items. The statistical properties of 
the model are consistent with observed changes 
in longitudinal datasets of proxies for cultural 
item popularity, such as steeper gradients around
popularity peaks, indicating faster rises and declines~\cite{attention_dynamics}.

%----------------------------------------
\subsection{Framework}
%----------------------------------------

To make the discussion concrete, 
we briefly outline the framework used in our study.
Here we extend~\cite{attention_dynamics}
and consider the behavior of $N$ competing cultural 
items, where individual activities $x_i$ are taken 
as a measure of popularity, such as the number of downloads 
for the case of songs and albums.
Our basic model is
\begin{equation}
    \begin{aligned} 
\label{eq:rescaled}
       T_x\,\dot{x}_i &= x_i \big[ 1 - (1 - w) h_i - 
w \langle x \rangle_i \big]\,, 
        \qquad\quad \langle x \rangle_i &= 
\frac{1}{N-1}\sum_{j\neq i} x_j\,, \\
    T_h\,\dot{h}_i &=  x_i -  h_i ,
    \end{aligned}
\end{equation}
where  $h_i$ is the history of item $i$.
The parameter $w\in[0,1]$ regulates the 
relative weights of self- and mutual interaction, 
with $T_x, T_h>0$ being the respective timescales.
One has $h_i\to x_i$ when the memory update is 
fast, that is, in the limit $T_h/T_x\to0$. The system 
reduces in this case to the standard  
Lotka-Volterra model \cite{gros2024complex};
see the discussion in Sect.~1.1 in the Suppl.~Mat.~\cite{supp}.

The differential equation for $\dot h_i$
in (\ref{eq:rescaled}) can 
be integrated expressively 
\cite{Wernecke2019, henrik2023mapping}, 
leading to the well-known representation
\begin{equation}
    \label{ref_h_i_exp}
    h_i(t) = \int_0^\infty d\tau\, K(\tau)\, x_i(t-\tau),
    \qquad\quad  K(\tau) = \frac{1}{T_h}\,\exp(-\tau/T_h)\,,
\end{equation}
where $K(\tau)$ is the memory kernel. State 
histories are thus encoded by $h_i(t)$ in 
terms of an exponential moving average.

%%%%%%%%%%%%%%%%%%%%%%%%%%%%%%%%%%%%%%%%%
\begin{figure} [th!] 
\centering
\includegraphics[width=0.6\textwidth]{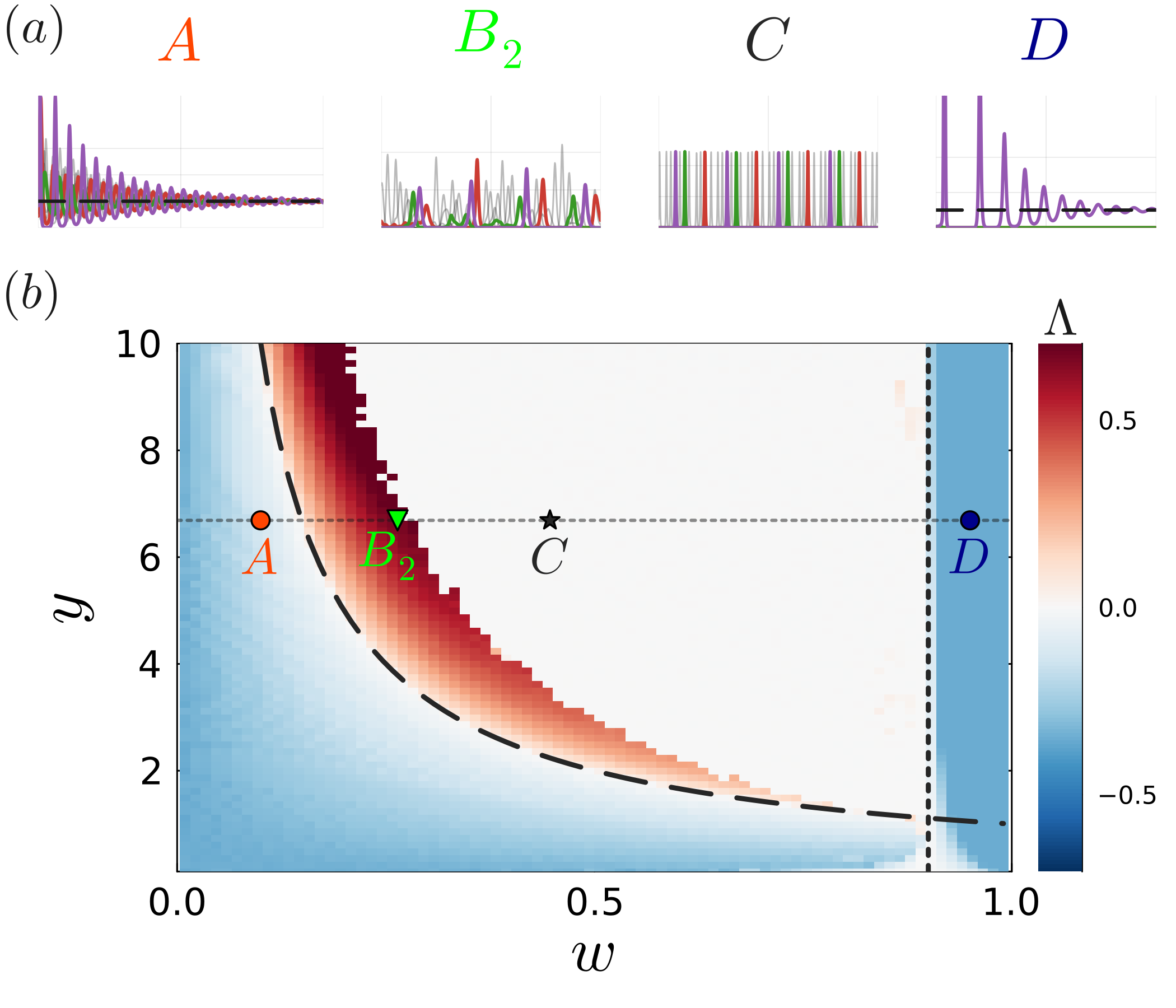}
\caption{\textbf{Phase diagram for $N=10$ items.}
Shown are results for the symmetric and homogeneous 
Lotka-Volterra with memory, as defined by
(\ref{eq:rescaled}) and (\ref{ref_h_i_exp}),
together with $T_h=1$. For details see
Sect.~\ref{sec:regimes} and Sect.~\ref{sect_numerics}.
\\
(a) Time series, illustrating the four possible
types of dynamics: convergence to the coexistence 
fixed point, chaotic winnerless competition (WLC), 
periodic WLC, and convergence to the winner-takes-all
(WTA) fixed point.
Parameters correspond to A, B$_2$, C, and D indicated
in the phase diagram below. For simplicity, only three 
activities are colored; the others are plotted in gray.\\
(b) The phase diagram, color-coded by the largest 
Lyapunov exponent, $\Lambda$. On the y-axis, a
rescaled timescale $y=1/(T_x(N-1))$ has been 
used, compare (\ref{eq_y_c}). The thick black dashed 
curve and the black dotted vertical line, defined by
Eqs.~(\ref{eq:coexistance_condition}) and 
(\ref{eq:WTA_condition}),
respectively, indicate the stability boundary 
of the coexistence and the WTA-type dominance 
fixed points, with blue shading corresponding 
to stable regimes. Specific parameter values
used as examples for time series are indicated by 
the markers and capital letters. 
    }
    \label{fig:phase_diagN10}
\end{figure}
%%%%%%%%%%%%%%%%%%%%%%%%%%%%%%%%%%%%%%%%%

%----------------------------------------
\subsection{Symmetric Lotka-Volterra models}
%----------------------------------------

In a symmetric all-to-all setting with $N$ competitors, 
where all coupling strengths are identical, as for
(\ref{eq:rescaled}), homogeneity restricts 
the long-term dynamics to two possible
states~\cite{Cohen1983}. 
The behavior is determined by the balance of
the two fundamental couplings:
\begin{itemize}
\item Self-interaction, $1-w$ in Eq.~(\ref{eq:rescaled}),
which arises, e.g., from resource limitation or 
a ``boringness'' effect.
\item Inter-competitor coupling, $w$ in 
Eq.~(\ref{eq:rescaled}).
\end{itemize}
When self-interaction dominates, population (or activity) 
levels can simultaneously remain at nonzero values,
as illustrated in
Fig.~\ref{fig:phase_diagN10}(a), panel A.
In contrast, when competition dominates, 
only a single competitor remains active, leading 
to a winner-takes-all (WTA) behavior
(compare Fig.~\ref{fig:phase_diagN10}(a), panel D). 
In both cases, convergence to the respective fixed 
points represents the only possible asymptotic outcomes.

LV-type systems can serve as a basis for attention 
dynamics models only if there exists a parameter regime 
that allows for winnerless competition (WLC), i.e., 
long-term irregular (chaotic) or regular
fluctuations in which the popularity of items 
alternates indefinitely. Examples are shown
in Fig.~\ref{fig:phase_diagN10}(a), 
panels B and C, respectively.

% %----------------------------------------
% \subsection{Alternative approaches}
% %----------------------------------------

% Let us put the framework used here in context.
% Within large interconnected systems, such 
% as competitive LV systems, multiple dynamical 
% phases can arise depending on the coupling
% structure and the strength of the interactions. 
% This is case for Eqs.~(\ref{eq:rescaled}) and 
% (\ref{ref_h_i_exp}), other venues are, however
% also possible.

% For example, complexity may be induced
% in large LV systems by asymmetric and/or disordered
% interactions~\cite{Bunin1,Bunin2,GenerLV}, 
% by the external environmental
% noise~\cite{Opper92}, or by 
% migration~\cite{Bunin1,Bunin3}.  Chaotic
% attractors in asymmetric systems \cite{Vano4D},
% as well as switching dynamics induced by 
% instabilities — where trajectories temporarily 
% visit saddle points connected by
% heteroclinic connections~\cite{Ashwin2005,Voit2019}
% -- have been reported in various contexts, 
% ranging from motor activities driven by
% chaotic neural networks~\cite{Varona2002} to animal search
% patterns~\cite{Gutirrez2015}.

% Other studies aimed to determine analytically the 
% conditions under which winnerless chaos (WLC) occurs, 
% such as in discrete LV variants, often assuming
% some form of special symmetry in the interaction
% matrix~\cite{GonzlezDaz2013}.

%----------------------------------------
\subsection{Structure of the paper}
%----------------------------------------

% Since the stability analysis of the model proposed
% by Lorenz-Spreen et al.~\cite{attention_dynamics} has only been
% carried out for the single-topic and two-competing-topic cases, 

We investigate the model defined by
(\ref{eq:rescaled}) and (\ref{ref_h_i_exp})
in detail, both analytically and numerically. 
The first objective is to determine the parameter 
regions in which winnerless chaos is possible
(Sect.~\ref{sec:WLC}). Once this is established, 
we examine the resulting dynamics by extending 
the definition of chart lifetimes, with the aim 
to study how lifetime distributions change 
with respect to the relevant control parameters
(Sect.~\ref{sec:charting}). Our main results
are presented in two places:
\begin{itemize}
\item Sect.~\ref{sect_max_chaos}:
Here the notion of maximally chaotic
cultural dynamics is discussed.
\item Sect.~\ref{sect_lifetime_stats}:
Here we show how cultural acceleration leads to 
power law chart lifetimes.
\end{itemize}
Further information is provided in the Appendix
and in the Supplementary Material \cite{supp}.

%%%%%%%%%%%%%%%%%%%%%%%%%%%%%%%%%%%%%%%%%%%%%%%%%%%%%
\section{Winnerless competition}
\label{sec:WLC}
%%%%%%%%%%%%%%%%%%%%%%%%%%%%%%%%%%%%%%%%%%%%%%%%%%%%%

It has been shown that delays can destabilize 
fixed points through Hopf bifurcations in LV 
systems and other dynamical 
systems~\cite{Wernecke2019,word_usage,Rahman2015}.
% In delayed LV systems, self-sustained oscillations can arise 
% when the interaction structure is 
% symmetric~\cite{Cohen1983} or non-symmetric \cite{Gopalsamy1980}. 
However, delay-induced destabilization and Hopf 
bifurcations in competitive Lotka--Volterra systems 
have previously been studied mostly in low-dimensional 
settings~\cite{song2004stability}. Here, we first show 
that symmetries lead to a large number of coexisting 
fixed points and limit cycles. We then carry out a 
complete analytical stability analysis to delineate 
the parameter regime of WLC dynamics.

%----------------------------------------
\subsection{Symmetries may induce multistability}
%----------------------------------------

Coupled identical systems can exhibit long 
transients due to the presence of a large number 
of coexisting attractors and an extended chaotic 
saddle in phase space~\cite{sandor2015versatile}. 
Multistability is induced both by the symmetry of 
the interactions and by the assumption that the 
parameters describing the isolated competitors are 
identical. Due to the symmetric coupling and the 
assumption that all competitors have the same growth 
rate, there exist symmetry operations $\bm{\sigma}_{ij}$ 
that act on the state vector 
$\vect{u}=(x_1,x_2,\dots,x_N,h_1,h_2,\dots,h_N)$ 
by interchanging the indices $i$ and $j$ simultaneously 
in both the $x$ and $h$ variables, while leaving the 
dynamics unchanged. Using the symmetry operators 
$\bm{\sigma}_{ij}$, different attractors can be 
transformed into one another 
(see Appendix~\ref{ap:symmetries}).

The effect of embedded symmetries is most apparent 
when the system exhibits periodic behavior. In 
this case, the high-activity peaks alternate 
periodically between the different competitors, 
each of them emerging as a temporary winner of the 
competition. We can differentiate visually between 
these cycles if we look at the order in which they 
follow each other in the time series, since two 
different orders cannot correspond to the same 
cycle by continuity of the phase space (solutions are unique).

For a system with $N$ competitors, there are $(N-1)!~ $ 
equivalent limit cycles, in terms of ordering, with
each attractor having its own basin of attraction.

%----------------------------------------
\subsection{Stationary solutions}
\label{sec:fixedPoints}
%----------------------------------------

Here we show that the conditions for winnerless
competition can be inferred from the linear stability 
of the fixed points
$$
\fixp{u} = (\fixp{x}, \fixp{h}) =
(\fixpval{x}[M][1],\fixpval{x}[M][2],...,
\fixpval{x}[M][N],\fixpval{h}[M][1],
\fixpval{h}[M][2],...,\fixpval{h}[M][N])
$$
of the delayed Lotka-Volterra system
(\ref{eq:rescaled}).
The stationarity condition
$\fixpval{x}[M][i]=\fixpval{h}[M][i]$ implies
that the fixed points are structurally identical 
to the fixed points of the classical LV system 
(see the Suppl.~Mat.~\cite{supp}).

A general fixed point will have $M\le N$ active
items, the winners, and $N-M$ inactive items.
The $\binom{N}{M}$ possible M-winner fixed 
points are structurally equivalent, as
a consequence of the permutation symmetry 
of (\ref{eq:rescaled}) and (\ref{ref_h_i_exp}).
Hence they can be constructed in terms of the 
unique permutations of the
elements of $\fixp[{M}]{x}$, 
\begin{equation} 
\label{eq:M_winner}
\fixp[{M}]{u} = (\fixp[{M}]{x},\fixp[{M}]{h}), \qquad \quad 
\fixp[{M}]{x} = \fixp[{M}]{h} =  ( 
\underbrace{\fixpval{x}[M],\fixpval{x}[M],...,\fixpval{x}[M]}_M,
\underbrace{0,0,...,0}_{N-M} )\,,
\end{equation}
where the nonzero activity value for $M>0$ 
can be evaluated explicitly as
(details are given in the Suppl.~Mat.~\cite{supp}),
\begin{equation}
\label{eq:fixedpoint}
\fixpval{x}[M] = \left\{
\begin{array}{ccl}
\frac{N - 1}{N-1 - w(N-M)} & \text{if}&
 N > 1\,, \\[1.0ex]
\frac{1}{1-w} & \text{if} &  N=1\,.
\end{array}\right.
\end{equation}
These results allow not only
to evaluate the Jacobian, but also to
factorize the resulting characteristic polynomial 
into four distinct terms \cite{supp}:
\begin{equation} 
\label{eq:charpoly2}
\underbrace{(-1-\lambda)^{N-M}}_{\lambda_a} ~ \cdot ~ 
\underbrace{(a_1-\lambda)^{N-M}}_{\lambda_b} ~ \cdot ~ 
\underbrace{\biggl[(-1-\lambda) (b_\lambda - a_2) \biggr]^{M-1}}_{\lambda_c^{\pm}} ~ \cdot ~ 
\underbrace{(-1-\lambda)(b_{\lambda}-a_3)}_{\lambda_d^{\pm}} = 0\,.
\end{equation}
with
$$
a_1 = \frac{1}{T_x}(1-\fixpval{x}[M] \cdot \frac{wM}{N-1}), \qquad 
a_2 = -\frac{1}{T_x}(w \cdot \frac{\fixpval{x}[M]}{N-1}),
$$
and
$$
b = (w-1) \cdot \fixpval{x}[M]/T_x, \qquad 
a_3 = -(M-1)a_2, \qquad 
b_{\lambda} = -\lambda - bc/(1 - \lambda)\,.
$$

%----------------------------------------
\subsection{Stability analysis}
\label{sec:stability}
%----------------------------------------

The four types of eigenvalues resulting from the 
characteristic polynomial (\ref{eq:charpoly2}) are
\begin{equation}  
\label{eq:eig}
\begin{array}{rclll}
\lambda_a &=& -1, &
m(\lambda_a) = N-M, & ~\text{if} ~~ M < N\,, \\[1.5ex]
\lambda_b &=& \frac{1}{T_x}(1-\frac{wM}{N-1} \cdot \fixpval{x}[M]),&
m(\lambda_b) = N-M, & ~\text{if} ~~ M < N\,, \\[1.5ex]
\lambda_c^{\pm} &=& \frac{1}{2} \left[-(1+a_2) \pm 
\sqrt{(1+a_2)^2 + 4(b - a_2)}\right], &
m(\lambda_c^{\pm}) = M-1, & 
~\text{if} ~~ M \geq 2\,, \\[1.5ex]
\lambda_d^{\pm} &=& \frac{1}{2} \left[-(1+a_3) \pm 
\sqrt{(1+a_3)^2 + 4(b - a_3)}\right], &
m(\lambda_d^{\pm}) = 1, & 
~\text{if} ~~ M \neq 0\,,
\end{array}
\end{equation}
where we denoted with $m(\lambda)$ the algebraic 
multiplicity of the eigenvalue.  
We will be primarily interested in how the dynamics changes
if the interaction parameter $w$ is changed, as it controls 
both competition and the self-saturation strength. 

The trivial fixed point of total extinction, 
$\fixp[0]{u}$, is always a  saddle, having $N$ 
positive and $N$ negative eigenvalues, respectively 
$\lambda_a$ and $\lambda_b$. For the non-trivial 
fixed points with $M>0$, however, the stability 
conditions in terms of each eigenvalue
type are given by:
\begin{align}
\lambda_a = -1 &< 0\,,\label{eq:eig_cond_a}\\
\lambda_b &< 0 
\qquad\Leftrightarrow\qquad 
w > \frac{N-1}{N}\,, \label{eq:eig_cond_b}\\
\Re(\lambda_c^{\pm}) &< 0 
\qquad\Leftrightarrow\qquad 
w < \frac{N-1}{1/T_x + N-M} 
\qquad\text{and}\qquad 
w < \frac{N-1}{N}\,, \label{eq:eig_cond_c}\\
\Re(\lambda_d^{\pm}) &< 0 
\qquad\Leftrightarrow\qquad 
w < \frac{N-1}{N-M}\,, \label{eq:eig_cond_d}
\end{align}
From Eq.~\eqref{eq:eig}, it can also be inferred 
that not every type of eigenvalue is present in 
the spectrum of each fixed point~$\fixp[{M}]{u}$. 
The composition of this spectrum, consisting
of $2N$ eigenvalues in total, depends on $M$.
A summary is given in Table~\ref{tb:eigspectra}.

%%%%%%%%%%%%%%%%%%%%%%%%%%%%%%%%%%%%%%%%%%%%%%%%%%%
\begin{table}[b] 
\caption{The algebraic multiplicity of the eigenvalue spectrum of
groups of fixed point $\fixp[{M}]{u}$. When the eigenvalue type is
present in the spectrum for a particular $M$, the multiplicity is
shown. 
%Note that both $\lambda_c^{\pm}$ and $\lambda_d^{\pm}$
%represent pairs, the multiplicity refers to both ($+$ and $-$)
%eigenvalues. 
The $\cross$ signs indicate cases in which a given
eigenvalue type does not exist. Rows add up to
$2N$, the total number of eigenvalues.} 
\label{tb:eigspectra}
\centering
\def\arraystretch{1.5}%
\begin{tabular}{|c|c|c|c|c|}
\hline
    $M$ & $\lambda_a$ & $\lambda_b$ & $\lambda_c^{\pm}$ & $\lambda_d^{\pm}$ \\
    \hline \hline
     $0$ & $N$ & $N$ & $\cross$ & $\cross$  \\
     \hline
     $1$ & $N-1$ & $N-1$ & $\cross$ & $1$  \\ 
     \hline
     $2,3,...,N-1$ & $N - M$ & $N - M$ & $M-1$ & $1$  \\
     \hline
     $N$ & $\cross$ & $\cross$ & $N-1$ & $1$  \\
     \hline
\end{tabular}
\end{table}
%%%%%%%%%%%%%%%%%%%%%%%%%%%%%%%%%%%%%%%%%%%%%%%%%%%

%----------------------------------------
\subsection{Dynamical regimes}
\label{sec:regimes}
%----------------------------------------

Based on the above conditions, one sees that 
M-winner fixed points with $M = 2,3,...,N-1$ 
are never stable. This is because the eigenvalue 
spectrum would contain in this case all four 
types, with the conditions (\ref{eq:eig_cond_b}) 
and (\ref{eq:eig_cond_c}) contradicting each other. 
Therefore, the possible
asymptotic dynamical regimes are:
\begin{itemize}
\item[--] \textbf{(A) Coexistence.} 
The unique $\vect{u}^*_\mathrm{c} \equiv \fixp[{N}]{u}$ 
fixed point ($M=N$, all identical activities) 
has two stability conditions, namely
\begin{equation}\label{eq:coexistance_condition}
 w<w_c\,, \qquad w_c = T_x(N-1), \qquad 
\mbox{if} ~~ 1>N T_x\,,
\end{equation}
and
\begin{equation}\label{eq:coexistance_condition2}
w<w_d\,, \qquad w_d = \frac{N-1}{N}, \qquad 
\mbox{if} ~~ 1<N T_x\,,
\end{equation}
see conditions~(\ref{eq:eig_cond_c}) and (\ref{eq:eig_cond_d}).

\item[--] \textbf{(B,C) Chaotic and periodic WLC}. 
For $1 / T_x > N$, there is an interval in which none 
of the fixed points is stable and winnerless dynamics 
emerges, namely when 
\begin{equation}\label{eq:WLC_condition}
    w_c<w<w_d\,,
\end{equation}
with the interval spanned by $w_c$ and $w_d$ 
reading graphically as
$$\includegraphics[width=0.4\linewidth]{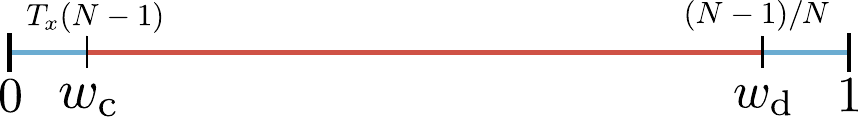}
$$
For $1/T_x < N$, no winnerless competition 
is possible since there is at least one stable 
fixed point for the entirety 
of the $w \in [0,1]$ interval, as now $w_d < w_c$.
WLC may be of two types: chaotic (B) or periodic (C). 

Dynamically, WLC emerges at $w_c$ via
a high-dimensional Hopf bifurcation (see Appendix \ref{ap:dynamics_details}), namely
at the point where the coexistence fixed point 
$\vect{u}^*_\mathrm{c}$ becomes unstable.
\item[--] \textbf{(D) WTA}. 
The $N$ equivalent winner-takes-all
fixed points 
$\vect{u}^*_\mathrm{d} \equiv \fixp[{1}]{u}$ 
are stable for
        \begin{equation}\label{eq:WTA_condition}
        w>w_d\,, 
        \end{equation}
see conditions~(\ref{eq:eig_cond_a}), (\ref{eq:eig_cond_b}), 
and (\ref{eq:eig_cond_d}).
\end{itemize}
Our naming of the different dynamical regimes
in terms of letters A, B, C, D has been used for 
the $N=10$ and $N=300$ phase diagrams presented
Figs.~\ref{fig:phase_diagN10} 
and \ref{fig:phase_diagN300}.
There, we did not use $(w,T_x,T_h)$ for encoding 
phase space, but $(w,y)$, with
\begin{equation}
\label{eq_y_c}
y= \frac{1}{T_x(N-1)}, \qquad
y_c = \frac{1}{w}\,,
\end{equation}
where the last relation is 
the critical line $w_c=T_x(N-1)$, see
(\ref{eq:coexistance_condition}).
In addition, we set $T_h=1$ in (\ref{eq:rescaled}) as our default.
With (\ref{eq_y_c}), the stability boundary 
(\ref{eq:coexistance_condition}) of the
coexistence regime collapses onto a single 
curve for all sizes $N$, with a shrinking WTA 
interval delimited by (\ref{eq:coexistance_condition2}),
as $N$ increases. 

In Appendix~\ref{ap:dynamics_details}, a detailed bifurcation
analysis as a function of the number of items $N$ is
presented. E.g., one finds that the degeneracy
resulting from the permutation symmetry of 
(\ref{eq:rescaled}) affects the unfolding of
the Hopf bifurcation, inducing at the same time extended 
chaotic transients in regions of the phase diagram
where the winnerless competition leads otherwise to 
stable limit cycles.

%----------------------------------------
\subsection{Numerics}
\label{sect_numerics}
%----------------------------------------

The above analytic results are confirmed by
numerical simulations. In summary:
\begin{itemize}
\item Below the Hopf curve, as defined by 
(\ref{eq:coexistance_condition}) in case
\textbf{(A)}, initial fluctuations decay exponentially 
due to a negative largest Lyapunov exponent, 
$\Lambda < 0$, leading in the end to equally distributed 
activities for all coupling strengths $w < w_c$. 
For $N=10$, an example is shown in panel A of 
Fig.~\ref{fig:phase_diagN10}(a). 

\item On the other hand, for large
coupling strengths, $w > w_d$ in case \textbf{(D)}, 
a single item receives all the attention, as shown
in  panel D of Fig.~\ref{fig:phase_diagN10}(a).

\item Between these two phases, the regime of winnerless competition, 
cases \textbf{(B)} and \textbf{(C)}, may lead to either
periodic oscillations or chaotic fluctuations, depending 
on the number of competitors and the exact parameter 
values; panels B and C
of Fig.~\ref{fig:phase_diagN10}(a). 
\end{itemize}
For $N=300$, the phase diagram is presented in
Fig.~\ref{fig:phase_diagN300}(a). For the discussion of the methods used for numerical integration, see Appendix \ref{ap:numerics}. With respect to $N=10$,
the region with winnerless chaos is greatly 
enlarged.\footnote{For $N=300$ we rescaled negative $\Lambda$ values by a
factor of $10$ for better visibility.}

%%%%%%%%%%%%%%%%%%%%%%%%%%%%%%%%%%%%%%%%%%%%%%%%%%%%%%%%%%%%%
\begin{figure} [t] 
\centering
\includegraphics[width=0.8\textwidth]{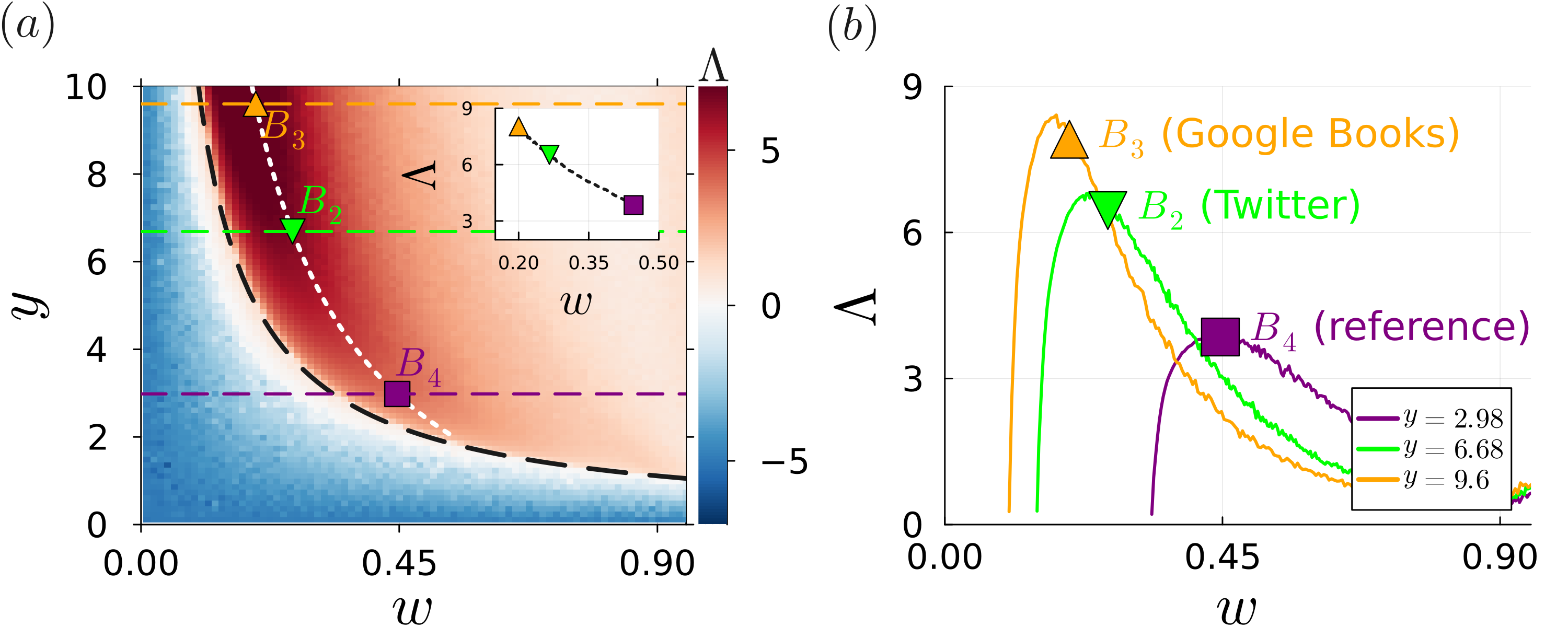}
\caption{\textbf{Phase diagram for $N=300$ items.} 
(a) As for Fig.~\ref{fig:phase_diagN10}. Note the
enlarged region with winnerless chaos, characterized
by positive largest Lyapunov exponents $\Lambda$ (in red).
The inset shows $\Lambda$ as a function of $w$ 
along the white dotted curve, defined by (\ref{eq:rcurve}). 
% Negative exponents with $\Lambda<0$ are scaled by a
% factor of $10$ for better visibility. \\
%
(b) For three different values of $y=1/(T_x(N-1))$, 
corresponding to the respective color-coded horizontal
lines in the phase diagram, the largest Lyapunov exponents are shown
as a function of $w$. The markers
$B_2$, $B_3$ and $B_4$  
are located close to the point where $\Lambda$ is largest.
The associated labels in terms of cultural item datasets
are discussed in Sect.~\ref{sect_max_chaos}.
}
\label{fig:phase_diagN300}
\end{figure}
%%%%%%%%%%%%%%%%%%%%%%%%%%%%%%%%%%%%%%%%%%%%%%%%%%%%%%%%%%%%%

%----------------------------------------
\subsection{Maximal Chaos}
\label{sect_max_chaos}
%----------------------------------------

In Fig.~\ref{fig:phase_diagN300}(b) the
evolution of the largest Lyapunov exponent 
$\Lambda$ as a function of $w$ is presented 
for $N=300$ when $T_x$ is kept constant, 
respectively $y=1/(T_x(N-1))$. Right after the bifurcation of the coexistence fixed point, viz when it becomes unstable for $w>w_c$,
the winnerless competition is chaotic (as indicated by the positive $\Lambda$ values). A 
pronounced peak is observed, with the position
depending on the value of $y$; shown are
$y=2.98$ ($B_4$), $y=6.68$ ($B_2$) and $y=9.6$ ($B_3$),
where the selected locations $B_{2/3/4}$ 
are marked in the phase diagram.

\begin{figure} [t] 
\centering
\includegraphics[width=0.8\textwidth]{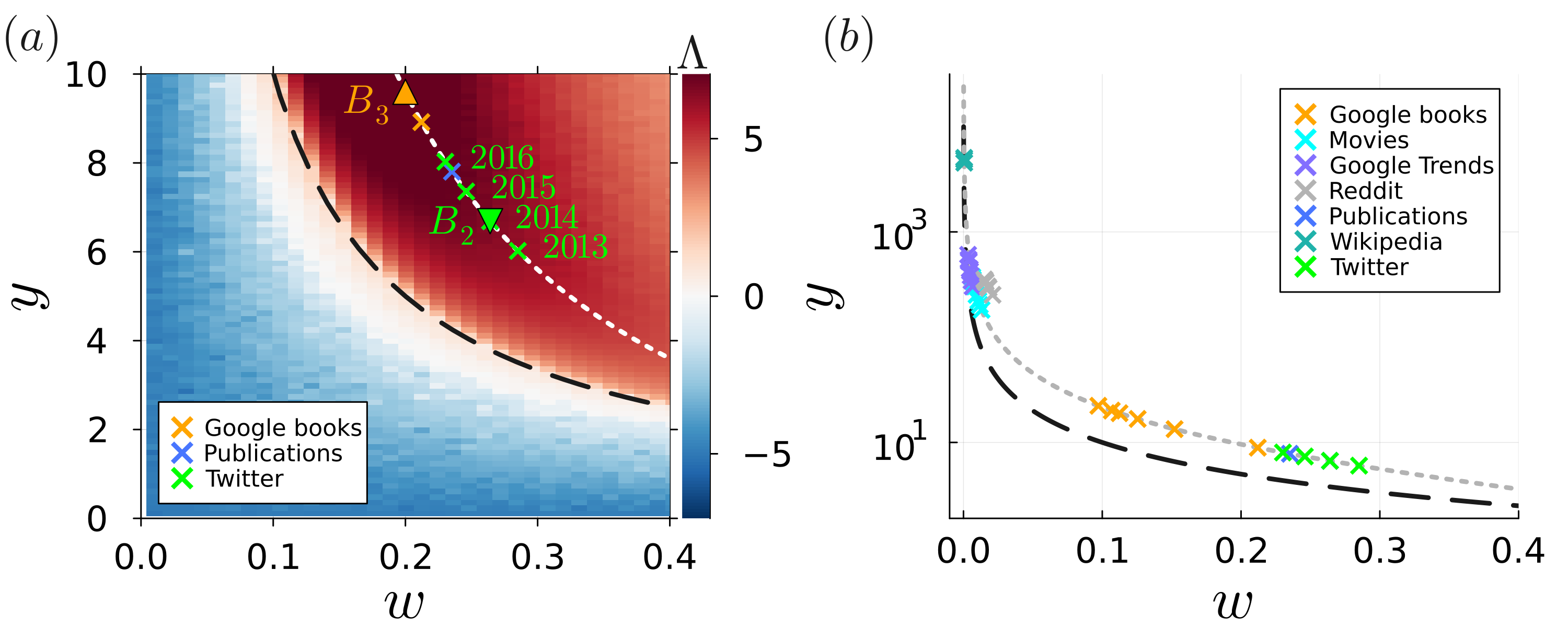}
\caption{\textbf{Optimal parameters for cultural datasets.} 
Optimal pairs of parameters $(w,y)$, as extracted from 
fitting $N=300$ simulations to various cultural item
datasets (from \cite{attention_dynamics}). The dotted
white/gray lines (left/right) denote (\ref{eq:rcurve}).
(a) As for Fig.~\ref{fig:phase_diagN300}, but for $0<w<0.4$.
Included are the optimal data pairs for Google Books, Twitter, and
scientific publications.
(b) As for (a), but with an enlarged $y$-axis (log scale). We dropped
the estimates of the largest Lyapunov exponents $\Lambda$, which
we did evaluate only for $y<10$.
}
\label{fig:phase_diagN300_datasets}
\end{figure}

In order to connect with the behavior of real-world
cultural items, we replotted the $N=300$ phase diagram
in Fig.~\ref{fig:phase_diagN300_datasets}(a) for $w<0.4$, including
this time the locations of the optimal pairs of parameters 
$(w,y)$ that have been obtained by the authors 
of \cite{attention_dynamics} by fitting $N=300$ simulations 
to several cultural item datasets. Most of the optimal $(w,y)$ pairs
are located on the curve 
\begin{equation} \label{eq:rcurve}
    y = c(1-w)/w \,,
\end{equation}
where $c$ is a constant (see the white dashed curve on Figs.~\ref{fig:phase_diagN300}(a) and~\ref{fig:phase_diagN300_datasets}(a)). The interpretation of the parameter $c$ is discussed in Sect.~\ref{sec:charting}.
\begin{itemize}
\item As shown in Figs.~\ref{fig:phase_diagN300}(a) and~\ref{fig:phase_diagN300_datasets}(a)
   the $B_{2/3/4}$ markers are located close to maximal chaos 
    in terms of Lyapunov exponents $\Lambda$. These three markers
    are on the white dotted line defined by (\ref{eq:rcurve}). Indeed,
    this line tracks max-chaos well for $y<10$, viz in the region
    where $\Lambda$ was evaluated.
\item In Fig.~\ref{fig:phase_diagN300_datasets}(b) a log-$y$ axis
    has been used in order to accommodate
    additional types of cultural items,  including Google Books and 
    Trends, Movies, Reddit, Wikipedia, Twitter, and scientific publications, see~\cite{attention_dynamics}.
\end{itemize}
All estimated parameter combinations are located to the 
right of the instability line $y_c=1/w$, as defined by 
(\ref{eq_y_c}), viz in the chaotic region.
At this point we did not evaluate the Lyapunov 
exponents for $y>10$. 
However, the phenomenology that can be inferred from
Fig.~\ref{fig:phase_diagN300}(b) suggests that 
the experimentally relevant $(w,y)$ combinations are
located close to maximal chaos also for $y>10$. These observations
offer the hypothesis that the competition for
attention among cultural items may lead in many
instances to maximally chaotic dynamics.

%%%%%%%%%%%%%%%%%%%%%%%%%%%%%%%%%%%%%%%%%%%%%%%%%%%%%%%%%%%%%
%%%%%%%%%%%%%%%%%%%%%%%%%%%%%%%%%%%%%%%%%%%%%%%%%%%%%%%%%%%%%

%%%%%%%%%%%%%%%%%%%%%%%%%%%%%%%%%%%%%%%%%
\section{Charting lifetime distributions}
\label{sec:charting}
%%%%%%%%%%%%%%%%%%%%%%%%%%%%%%%%%%%%%%%%%

Popularity dynamics in sociocultural systems 
is characterized by a continuously fluctuating 
winnerless competition that can be measured
via different proxies for the popularity 
of cultural items (e.g., the number of likes 
per unit of time)~\cite{Schneider2021}. Our analysis
is based on the equivalence between 
the model originally introduced in~\cite{attention_dynamics},
\begin{equation}
\label{eq:delay_split_orig} 
\dot{L}_i = r_p L_i \left( 1 - \frac{r_c}{K}Y_i - 
c \sum_{j\neq i}^N L_j \right), \qquad\quad
\dot{Y}_i = L_i - \alpha Y_i\,,
\end{equation}
and the formulation used here, Eq.~(\ref{eq:rescaled}).

In (\ref{eq:delay_split_orig}), item 
activities are $L_i$, with the
parameters $r_\mathrm{p}$ and $r_\mathrm{c}$ 
denoting the production and consumption rates, 
$K$ the carrying capacity and
$c$ the coupling strength known from classical
population growth models with competition. The 
characteristic timescale is given by $1/\alpha$.
The delay term reduces to the instantaneous 
value of $L_i$ in the $\alpha \rightarrow \infty$
limit, and to the equally-weighted history case 
in the limit $\alpha \rightarrow 0$. 

To prove the equivalence of \eqref{eq:delay_split_orig} 
with (\ref{eq:rescaled}) one rescales the 
corresponding variables,
$$
\begin{array}{rclrcl}
x_i &=& L_i/\eta_L,\qquad
\eta_L &=& \alpha K/(r_c + \alpha Kc(N-1))\,, \\[1ex]
h_i &=& Y_i/\eta_Y,\qquad
\eta_Y &=& K/(r_c + \alpha Kc(N-1))\,.
\end{array}
$$
In addition one needs to rescale time $t = \alpha \cdot t_L$, 
where we denoted time in the $(x,h)$ system  (\ref{eq:rescaled})
as $t$, respectively as $t_L$ in the $(L,Y)$ formulation 
\eqref{eq:delay_split_orig}. 
With $T_h=1$, the relative coupling strength $w\in[0,1]$ 
is given together with the timescale $T_x$ 
by
\begin{equation}
\label{eq_w_y_T_x}
w = \frac{1}{1 + r/(\alpha c(N-1))}
  = \frac{1}{1 + y/c}, 
\qquad y = \frac{1}{T_x(N-1)},
\qquad T_x = \frac{\alpha}{r}\,,
\end{equation}
where we used $r=r_p=r_c$ and $K=1$,
as in \cite{attention_dynamics}.
Eq.~(\ref{eq_w_y_T_x}) with a constant coupling $c$ 
hence reduces to the parameter curve defined by (\ref{eq:rcurve}) 
in Sect.~\ref{sect_max_chaos}.

Besides the formal equivalence of Eqs.~(\ref{eq:rescaled}) 
and (\ref{eq:delay_split_orig}),
we note that (\ref{eq:rescaled}) 
represents a slow-fast formulation,
which implies that it is 
numerically stiff. For large-system numerical 
simulations, we therefore resorted
to \eqref{eq:delay_split_orig}, see also Appendix \ref{ap:numerics}.
As in \cite{attention_dynamics}, we use in
the following $r = r_p = r_c$, $K=1$ and $c=2.4$.
For this setting, only a single free parameter 
remains when the number of items $N$ is kept 
constant, namely $T_x=\alpha/r$.
 
%%%%%%%%%%%%%%%%%%%%%%%%%%%%%%%%%%%%%%%%%
\begin{figure} [t]
\centering
\includegraphics[width=0.7\textwidth]{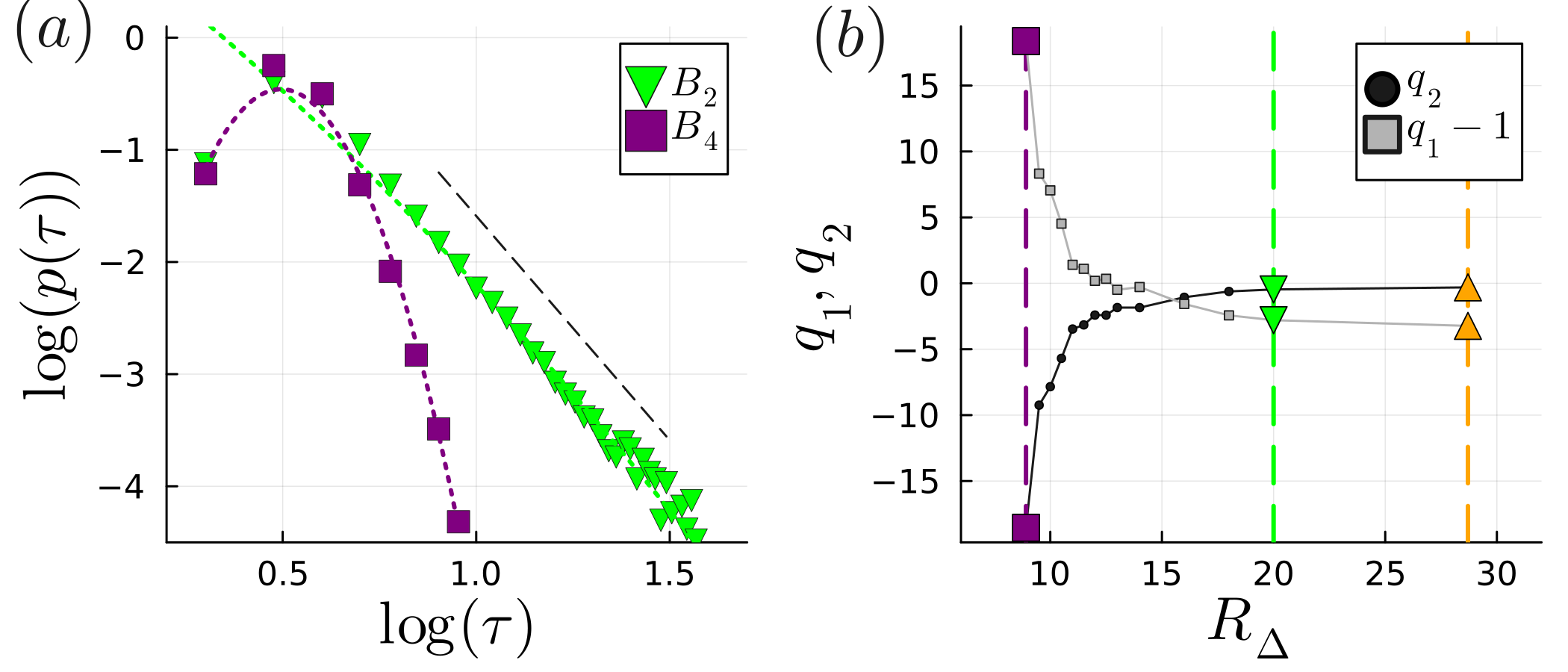}
\caption{\textbf{Transition from log-normal to power-law chart-lifetime distributions.}
For $N=300$ items, the chart lifetime distribution $p(\tau)$
resulting from (\ref{eq:popularity}).
(a) As a log-log plot, for the $B_2$ and $B_4$ sets of parameters,
see Fig.~\ref{fig:phase_diagN300}. Included are 
parabolic fits in log-space (dotted lines, see 
(\ref{eq_p_log_normal})), viz of the form 
$q_0 + (q_1-1)\log\tau + q_2(\log\tau)^2$. This
functional form corresponds to a log-normal $p(\tau)$
when $q_2\neq0$ and to a power law when $q_2\to0$. As a guide to the
eye, a power law decay $\propto \tau^{-4}$ has been added 
(gray dashed line).
(b) The quadratic and linear coefficients, $q_2$ and 
$q_1-1$, respectively, as a function of the
ratio $R_\Delta = \Delta t / T_x = r \cdot \Delta t_L$, where
$\Delta t$ is the charting period and $T_x$ the timescale of
cultural dynamics, as obtained when changing $w$ along the
line defined by (\ref{eq:rcurve}) 
or equivalently (\ref{eq_w_y_T_x}), compare
Fig.~\ref{fig:phase_diagN300}.
For fixed charting periods $\Delta t$, one observes 
a shift from variance-sensitive (log-normal) to mean-dominated
(power law) regimes when cultural dynamics accelerates,
viz when $T_x$ becomes smaller.
}
\label{fig:lifetimes_phasetransition}
\end{figure}
%%%%%%%%%%%%%%%%%%%%%%%%%%%%%%%%%%%%%%%%%

%----------------------------------------
\subsection{Defining charts and chart lifetimes}
%----------------------------------------
 
To bridge the gap between microscopic dynamics 
and observable macroscopic trends, we 
define a popularity metric based on integrated 
activities. This approach presumes that the 
activities $L_i(t_L)$ in (\ref{eq:delay_split_orig}) 
serve as proxies for instantaneous attention, representing 
metrics such as song plays or social media 
mentions~\cite{attention_dynamics}. Our aim
is to convert continuous item trajectories
into popularity charts, making use of the fact that
charts are based typically on aggregated measures, 
like the number of downloads per day, week 
or month. Hence we define the overall popularity
$\tilde{L}_i(t_L,\Delta t_L)$ of 
item $i$ over a trailing observation window $\Delta t_L$ as
\begin{equation}
\tilde{L}_i(t_L,\Delta t_L)=\int_{t_L-\Delta t_L}^{t_{L}}
L_i(\theta)~\mathrm{d}\theta\,.
\label{eq:popularity}
\end{equation}
In analogy with commercial cultural charts, we construct 
top-$k$ lists consisting of the $k$ items with the 
largest $\tilde{L}_i$ (a visual representation of this
procedure is given in Sect.~2 of the Supp. Mat.~\cite{supp}). 

Of particular interest are the individual chart lifetimes
$\tau_i=0,1,2,\dots$, which are defined as the number of consecutive 
windows of length $\Delta t_L$ during which an item
remains in a top-$k$ chart. The respective lifetime 
statistics $p(\tau)$ describes how often a particular 
lifetime $\tau=0,1,2,..$ is present during the overall observation period,
viz when averaged over all items and the length of the simulation.

%----------------------------------------
\subsection{Charting periods in terms of dynamical timescales}
\label{sect_Delta_T_x}
%----------------------------------------
 
For the study of the lifetime distributions $p(\tau)$ 
we use the delayed Lotka--Volterra system defined by
(\ref{eq:delay_split_orig}), together with the
parameter values obtained by fitting statistical properties
of the evolution of item popularity, $L_i(t_L)$, to real data,
as done by the authors of \cite{attention_dynamics} for 
$N=300$ topics.  

Specifically, we consider the chart lifetimes
along the curve (\ref{eq_w_y_T_x}),
which lies inside the chaotic region, compare
Fig.~\ref{fig:phase_diagN300}(a). As shown in
Fig.~\ref{fig:phase_diagN300_datasets}, this manifold 
in the space of parameters is relevant for fitting simulations
with observations. A possible parameterization is 
via the overall rate $r=\alpha/T_x$, for $N=300$ and
with  constant $c = 2.4$, $K = 1.0$, $\alpha = 0.005$ 
in (\ref{eq:delay_split_orig}).

The time interval $\Delta t_L=2$ used for 
(\ref{eq:popularity}) is chosen such that
the range of possible lifetimes $\tau$, in 
particular their minimum and maximum values, 
is comparable to that observed in Billboard
charts~\cite{Schneider2021}. We note, importantly,
that the charting interval $\Delta t_L$ used when simulating 
(\ref{eq:delay_split_orig}) transforms via
\begin{equation}
\label{eq_Delta_t}
\Delta t = \alpha \cdot \Delta t_L, \qquad\quad
R_\Delta = \frac{\Delta t}{T_x} = r \cdot \Delta t_L,
\end{equation}
to an equivalent interval $\Delta t$ for
(\ref{eq:rescaled}), given the corresponding
rescaling of time. Using $r=\alpha/T_x$, we defined
in (\ref{eq_Delta_t}) the ratio $R_\Delta$, which
measures the charting interval $\Delta t$ in terms
of the timescale $T_x$ of item dynamics. The access
to this quantity, $R_\Delta$, allows us to test a central
hypothesis in the field of cultural dynamics, namely,
the question whether cultural-item statistics depend
qualitatively on the ratio of exactly these two timescales, 
viz the observation period and the scale of the underlying
item dynamics~\cite{Schneider2019,Schneider2021}.

%----------------------------------------
\subsection{Cultural acceleration leads to power-law chart lifetimes}
\label{sect_lifetime_stats}
%----------------------------------------

Our results for chart lifetime distributions 
along (\ref{eq_w_y_T_x}),
$y = 1/(T_x (N-1))=r/(\alpha(N-1))$,
are presented in 
Fig.~\ref{fig:lifetimes_phasetransition}.
Following~\cite{Schneider2021}, we analyze 
chart lifetimes in terms of a log-normal
distribution,
\begin{equation}
\label{eq_p_log_normal}
p(\tau) \sim \frac{1}{\tau}\,\exp(q_0 + q_1\log\tau + q_2(\log\tau)^2)\,, 
\end{equation}
which reduces to a power law $\tau^{q_1-1}$ when the 
quadratic term $q_2$ vanishes (for more
details see Appendix~\ref{ap:lifetimes} and
Section 2 of the Supplementary Material).
We find two qualitatively distinct behaviors when
changing $r$ or $y$, respectively $R_\Delta$:
\begin{itemize}

\item \textbf{$R_\Delta$ small.}\\
The distribution has a substantial log-normal 
contribution $q_2$, as for B$_4$ in
Fig.~\ref{fig:lifetimes_phasetransition}(a), which translates to
a quadratic function in log--log space. This behavior tracks the 
lifetime distributions of albums in the Billboard charts 
before the 1990s~\cite{Schneider2021}, when cultural dynamics
took longer to evolve than the weekly charting period.

\item \textbf{$R_\Delta$ large.}\\
In this case, lifetime distributions follow a
relatively steep power-law decay, $\sim\tau^{-4}$,
see B$_2$
in Fig.~\ref{fig:lifetimes_phasetransition}(a). This
type of behavior is in agreement with
empirical observations for recent Billboard and 
Spotify weekly charts~\cite{Schneider2021}, given that 
today's attention dynamics plays out within days, not weeks.
\end{itemize}
The overall evolution of the quadratic $q_2$ and linear 
$q_1-1$ terms as a function of $R_\Delta$ is
presented in Fig.~\ref{fig:lifetimes_phasetransition}(b).
The magnitude of $q_2$ decreases monotonically while 
$q_1-1\rightarrow -4$ for large $R_\Delta$.
Given that we kept
the charting period $\Delta t$ fixed, the transition from 
log-normally distributed chart lifetimes to a power law
takes place when $T_x$ becomes small, see (\ref{eq_Delta_t}),
viz when cultural dynamics accelerates.

%%%%%%%%%%%%%%%%%%%%%%%%%%%%%%%%%%%%%%%%%%%%%%%%%%%%%%%%%%%%%
\begin{figure} [t] 
\centering
\includegraphics[width=0.95\textwidth]{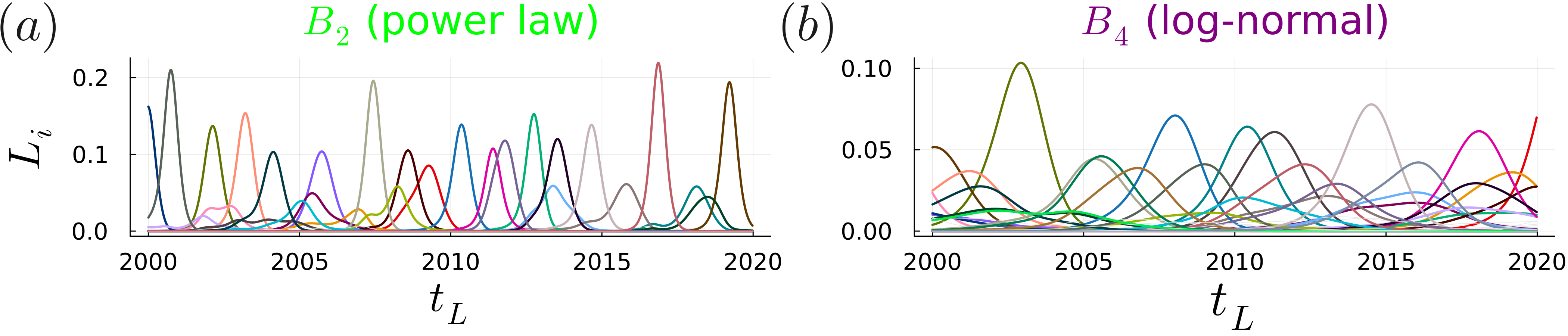}
\caption{\textbf{Item popularity timelines.} For two $N=300$ simulations of
(\ref{eq:delay_split_orig}), item activities as a function of time $t_L$
(color-coded). The two sets of parameters, $B_2$ (panel (a), $r=10$) and 
$B_4$ (panel (b), $r=4.46$),
correspond respectively to a power law and to a log-normal chart lifetime
distribution, as shown in Fig.~\ref{fig:lifetimes_phasetransition}.
At any time, only a limited subset of the $N=300$ item activities 
are large enough to be seen at scale. The shape of attention peaks tends 
to be more variable when $r$ is large, see discussion in
Sect.~\ref{sect_microscopic_analysis}.
}
\label{fig:timeseries_B2B4}
\end{figure}
%%%%%%%%%%%%%%%%%%%%%%%%%%%%%%%%%%%%%%%%%%%%%%%%%%%%%%%%%%%%%

%----------------------------------------
\subsection{Microscopic analysis}
\label{sect_microscopic_analysis}
%----------------------------------------

The results presented in 
Fig.~\ref{fig:lifetimes_phasetransition}
can be interpreted as a
consequence of a qualitative change in the structure of collective
attention allocation. For a microscopic analysis, we present
in Fig.~\ref{fig:timeseries_B2B4} the timeline of item
activities for parameter settings corresponding either
to a log-normal chart lifetime distribution ($B_4$, $r=4.46$), or
to a power law ($B_2$, $r=10$).

\begin{itemize}
\item \textbf{Small attention turnover rates $r$.} When 
attention dynamics is slow 
(B$_4$ in Fig.~\ref{fig:timeseries_B2B4}), activities are 
distributed across multiple items, allowing several topics to
simultaneously achieve comparable popularity levels. Activity
peaks tend to have similar shapes, which results in a 
comparatively high turnover rate of chart 
entries and a more homogeneous distribution 
of lifetimes, consistent with log-normal
statistics.

\item \textbf{Large attention turnover rates $r$.} For
a large production and consumption rate $r$
(B$_2$ in Fig.~\ref{fig:timeseries_B2B4}), 
activity peaks become increasingly sharp and
short-lived, leading to strong winner-dominated 
episodes. In this regime, only a small number of
topics capture the majority of attention, with
most other items remaining suppressed by at least 
an order of magnitude. There is less averaging among
different items, which leads to more variable
activity shapes. Some items may hence remain substantially
longer in the charts than others, which gives rise to 
heavy-tailed lifetime distributions.
\end{itemize}

This mechanism is consistent with the empirical findings of
Schneider et al.~\cite{Schneider2021}, where a similar 
transition from log-normal to power law lifetime
distributions in cultural charts over recent decades
was reported. The authors of~\cite{Schneider2021}
linked this crossover to a reduction in effective
decision times in consumption-based systems, driven by faster
information access and accelerated cultural turnover. 
In this interpretation, an analogous effect is achieved
by increasing the production and consumption rates $r$ in 
the Lotka-Volterra model (\ref{eq:delay_split_orig}), by
effectively reducing the timescale over which comparative 
evaluation occurs and thereby shifting the system
from variance-sensitive (log-normal) to mean-dominated 
(power law) regimes.

%%%%%%%%%%%%%%%%%%%%
\section{Discussion}
%%%%%%%%%%%%%%%%%%%%

The system~(\ref{eq:rescaled}) introduced here assumes homogeneous growth rates, identical delays, and a uniform all-to-all coupling scheme between competing items, without external effects or stochastic terms. Although real systems are inherently heterogeneous, these simplifying assumptions ensure that the model remains analytically tractable. The question of how variability in the parameters or a more realistic network topology between topics changes the dynamical landscape and the shape of lifetime distributions remains an open question for future research. 

The lifetime distributions in Fig.~\ref{fig:lifetimes_phasetransition} demonstrate that even continuous models with a macroscopic description, alongside agent-based microscopic approaches~\cite{Notarmuzi2018}, can produce interpretable results. It should be noted, however, that power laws~\cite{Weng2012} and log-normal distributions~\cite{Schneider2019} in the lifetimes of cultural items are common, but not universal~\cite{markovic2014power}.
The symmetry and homogeneity of the system lead to similar peak widths in the activities of different items, resulting in a characteristic lifetime when the charting interval is well below this timescale (see Sect.~2 of Suppl. Mat.~\cite{supp} for details).

As a future direction, one could also study alternative possibilities for the delay term; for example, the exponential kernel in the distributed delay could be replaced by a biexponential form~\cite{Candia2019}.
Another possibility is to make the delay vary with time, depending on the activities themselves. As demonstrated in~\cite{attention_dynamics}, the acceleration of engagement with popular content corresponds to an increase in the cultural yield $r$ within this framework. An accelerated attention economy implies that a topic's popularity reaches its peak earlier on average, in other words, shorter collective attention spans. However, this does not affect the characteristic delay time $1/\alpha$, suggesting that the timescale of collective memory remains constant. There is also a possible scenario in which collective memory is tied to a fixed number of recent events rather than a fixed time interval $1/\alpha$. In this case, an increased production rate, i.e., more activity peaks per unit time, would necessarily imply a shorter effective memory timespan. In other words, more rapid communication would also strengthen the effect of recency bias. This could be modeled by introducing a dependence $\alpha(r)$.

%%%%%%%%%%%%%%%%%%%%%
\section{Conclusions}
%%%%%%%%%%%%%%%%%%%%%

Delay-induced instability in competitive systems provides a minimal mechanism for generating complex, irregular attention dynamics without requiring heterogeneity or stochasticity. 
By combining analytical stability analysis with numerical simulations, we identified the parameter regimes in which sustained fluctuating dynamics can occur, even in a fully symmetric and homogeneous system.
We showed that, in contrast to classical Lotka-Volterra systems without delay, the introduction of memory effects destabilizes both coexistence and dominance fixed points over a finite parameter interval. This gives rise to a regime of winnerless competition, bounded by analytically derived stability thresholds. Within this regime, the system exhibits both periodic and chaotic dynamics, with chaotic fluctuations dominating for larger system sizes.

Building on this dynamical framework, we introduced a method to 
extract macroscopic observables—namely, chart lifetimes—from 
the microscopic activity dynamics. The resulting lifetime 
distributions reproduce qualitatively different types of 
behaviors, in accordance with empirical data for charts of
cultural items. In particular, we found a transition from 
log-normal to power law-like distributions as the production 
and consumption rates increase, capturing the acceleration 
of collective attention dynamics reported in real-world 
cultural systems. By evaluating the Lyapunov exponents, we also find that real sociocultural systems operate consistently near the point 
where the system is maximally chaotic and therefore maximally unpredictable.
\\

\textbf{Acknowledgments}
This work was supported by the grant of the Romanian Ministry of Research, Innovation and Digitization, CNCS-UEFISCDI, Projects No. PN-IV-P2-2.1-TE-2023-1548 and PN-IV-P6-6.1-CoEx-2024-0139.

\textbf{Code availability} Source code, written in the Julia programming language, for reproducing results and generating figures is available on GitHub, see Ref.~\cite{github}.

\clearpage
\printbibliography

%Appendix
\clearpage
\appendix

%%%%%%%%%%%%%%%%%%%%%%%%%%%%%%%%%%%%%%%%%%%%%%%%%%%%%%%%%%
\section{Conditions for winnerless competition with delay} 
%%%%%%%%%%%%%%%%%%%%%%%%%%%%%%%%%%%%%%%%%%%%%%%%%%%%%%%%%%

To avoid ambiguity, we briefly summarize the notation used throughout this section. We use $x_i(t)$ for the normalized activity and $h_i(t)$ for the corresponding history variable. The parameter $w \in [0,1]$ controls the relative strength of competition versus self-saturation (memory), whereas $T_x$ sets the effective time-scale separation between activity and memory dynamics. Averages over competitors are written as $\langle x \rangle_i$. 
When referring to fixed points, we use $M$ to denote the number of active (nonzero) components.

%----------------------------------------
\subsection{Symmetries}
\label{ap:symmetries}
%----------------------------------------

Due to the symmetric coupling in Eq.~(\ref{eq:rescaled}), there exist symmetry operations $\bm{\sigma}_{ij}$ acting on the state vector $\vect{u}$ under which the dynamics is invariant. More precisely, if a trajectory $\vect{u}(t)$ is a solution to the equations,
\begin{equation*}
    \dv{}{t} \vect{u}(t)  = \vect{F}(\vect{u}(t))\,,
\end{equation*}
then so is the trajectory that we get after applying the symmetry operation:
\begin{equation*}
     \dv{}{t} \bm{\sigma}_{ij} \vect{u}(t)  =
\vect{F}(\bm{\sigma}_{ij} \vect{u}(t))\,.  \end{equation*}

\begin{figure} [t] 
    \centering
    \includegraphics[width=0.8\textwidth]{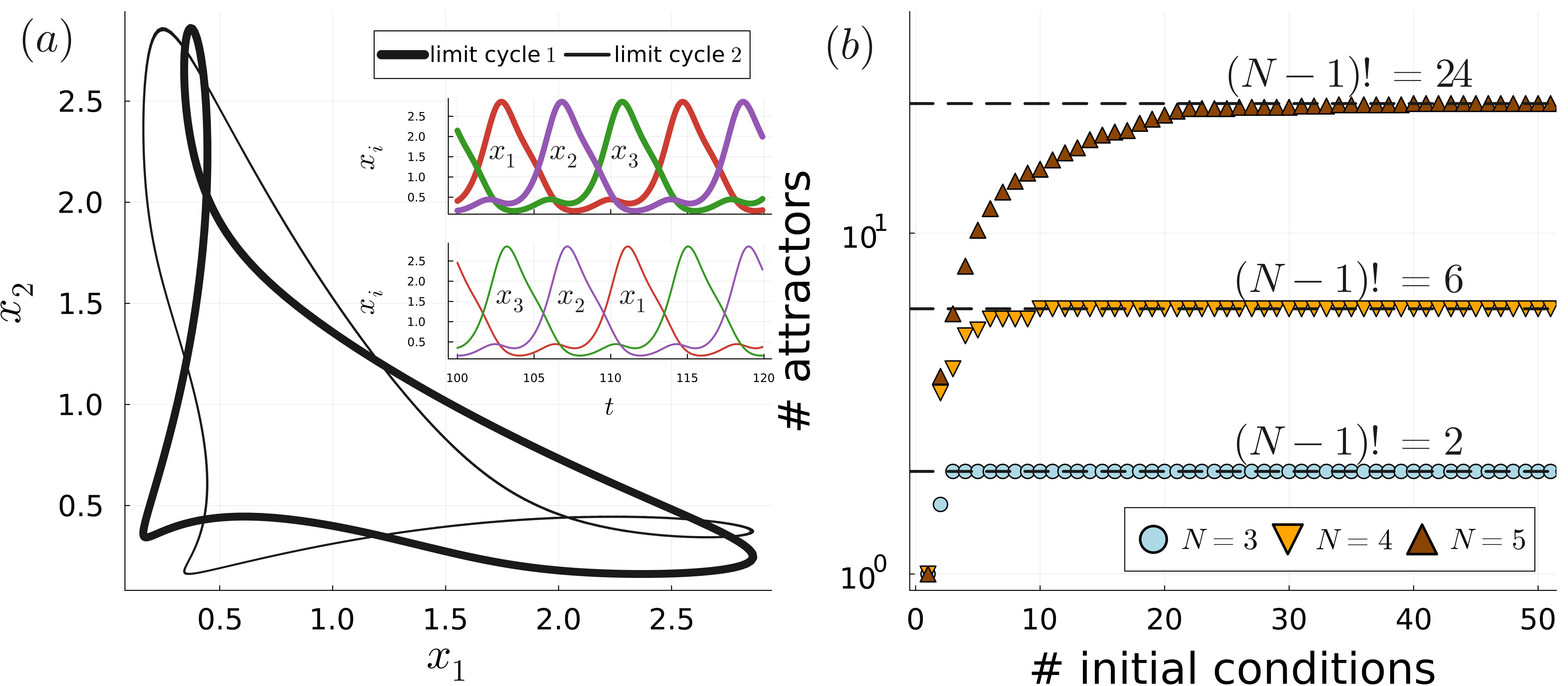}
    \caption{Symmetry-induced multistability. (a) Two
symmetry-related limit cycles (black and gray) in the case of $N=3$
competing topics projected to the $(x_1,x_2)$ plane for $w=0.6$ and
$1/T_x=3.5$. The inset shows the activation orderings in the time
series for the two limit cycles, respectively. (b) The number of
limit cycle attractors detected numerically using a method based on
recurrences in the partitioned phase space \cite{datseris2023} as a
function of the number of initial conditions used in the search
algorithm.  The limit cycles were found for three different systems
sizes $N=3/4/5$, marked with different shapes as given in the
legend. The theoretical limit $(N-1)!$ factorial is shown by the
horizontal dashed lines for the three different sizes, respectively.
The parameters used for the three systems are $w \in
\{0.6,0.6,0.7\}$ respectively, with $1/(T_x(N-1))=1.75$ for all
systems.} \label{fig:cycles}
\end{figure}

In general, $\bm{\sigma}_{ij}$ swaps the indices of the $x_i,h_i$, and $x_j,h_j$ variables, which is equivalent to swapping the order of two equations. 
These are related to elementary row-permutation matrices, which can be constructed by interchanging the corresponding rows of the identity matrix, but have two pairs of swapped rows, since we have to apply the interchange to the $h$ rows as well; hence the block form. 
Embedded symmetry is most apparent when the system exhibits periodic
behavior (limit cycles). In this case, the dynamics alternates
periodically between the different competitors, each of them
emerging as a temporary winner of the competition. We can
differentiate visually between these cycles if we look at the order
in which they follow each other in the time series (see
Fig.~\ref{fig:cycles}(a)), since two different orders cannot
correspond to the same cycle by continuity of the phase space
(uniqueness of solutions).  For a system with $N$ competitors, there are
$(N-1)!~ $ different limit cycles (orders), each attractor having
its own basin of attraction in phase space (see
Fig.~\ref{fig:cycles}(b)).

%%%%%%%%%%%%%%%%%%%%%%%%%%%%%%%%%%%%%%%%%
%%%%%%%%%%%%%%%%%%%%%%%%%%%%%%%%%%%%%%%%%

For example, in the $N = 3$ case, there are $3$ possible swaps:
\begin{equation}
    \bm{\sigma}_{12} = \begin{bmatrix} 
        0 & 1 & 0 & 0 & 0 & 0 \\ 
        1 & 0 & 0 & 0 & 0 & 0 \\
        0 & 0 & 1 & 0 & 0 & 0 \\
        0 & 0 & 0 & 0 & 1 & 0 \\
        0 & 0 & 0 & 1 & 0 & 0 \\
        0 & 0 & 0 & 0 & 0 & 1
        \end{bmatrix}\,, \quad
        \bm{\sigma}_{13} = \begin{bmatrix} 
        0 & 0 & 1 & 0 & 0 & 0 \\ 
        0 & 1 & 0 & 0 & 0 & 0 \\
        1 & 0 & 0 & 0 & 0 & 0 \\
        0 & 0 & 0 & 0 & 0 & 1 \\
        0 & 0 & 0 & 0 & 1 & 0 \\
        0 & 0 & 0 & 1 & 0 & 0
        \end{bmatrix}\,, \quad
        \bm{\sigma}_{23} = \begin{bmatrix} 
        1 & 0 & 0 & 0 & 0 & 0 \\ 
        0 & 0 & 1 & 0 & 0 & 0 \\
        0 & 1 & 0 & 0 & 0 & 0 \\
        0 & 0 & 0 & 1 & 0 & 0 \\
        0 & 0 & 0 & 0 & 0 & 1 \\
        0 & 0 & 0 & 0 & 1 & 0
        \end{bmatrix}
\end{equation}
and hence there are two cycles that can be transformed into one another (see Fig.~\ref{fig:cycles}(a)), namely by interchanging $x_1,x_2$ (respectively $h_1,h_2$) or $x_2,x_3$ (respectively $h_2,h_3$). Applying the symmetry operator $\bm{\sigma}_{13}$ would not give another cycle because the resulting order would be a cyclic permutation of the initial order. More precisely, if $\vect{u}_1(t)$ and $\vect{u}_2(t)$ are two distinct limit cycle solutions (thick black and thin gray curves in Fig.~\ref{fig:cycles}, respectively), then all the transformations
\begin{equation}
    \bm{\sigma}_{12} \cdot \vect{u}_1(t) = \vect{u}_2(t), \quad 
    \bm{\sigma}_{13} \cdot \vect{u}_1(t) = \vect{u}_2(t), \quad 
    \bm{\sigma}_{23} \cdot \vect{u}_1(t) = \vect{u}_2(t). 
\end{equation}
transform $\vect{u}_1(t)$ into $\vect{u}_2(t)$.

%%%%%%%%%%%%%%%%%%%%%%%%%%%%%%%%%%%%%%%%%
\begin{figure} [hbt!]
    \centering
    \includegraphics[width=\textwidth]{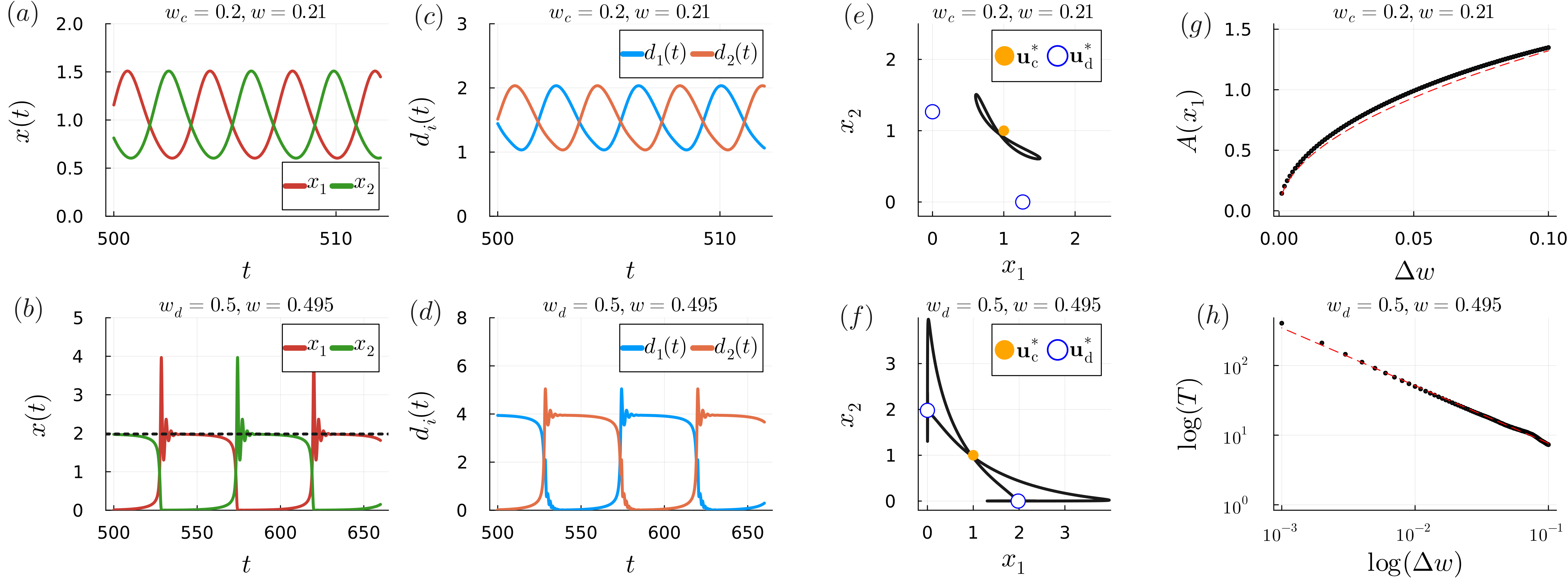}
    \caption{Dynamics close to the bifurcation points for the $N=2$ system with $1/T_x = 5$.
    \textit{Top row:} Vicinity of the Hopf bifurcation, $w = 0.21 > w_\mathrm{c} = 0.2$.
    \textit{Bottom row:} Approach to the heteroclinic bifurcation, $w = 0.495 < w_\mathrm{d} = 0.5$.\\
    (a,b) Time series of the activities $x_i(t)$; (c,d)
corresponding phase-space distances $d_i(t) = |\vect{u}(t) -
\fixp{u}{}_{\mathrm{d}_i}|$ from the two dominance fixed points,
$\fixp{u}{}_{\mathrm{d}_i} = \fixp[{1}]{u_i}$.  (e,f) Phase
portraits in the $(x_1, x_2)$ plane, showing the limit cycles
together with the unstable coexistence $\vect{u}^*_\mathrm{c}$ and
dominance $\vect{u}^*_\mathrm{d}$ fixed points (indicated in the
legends).
    (g) Amplitude scaling of the limit cycle, $A(x_1) = (\max(x_1) -
\min(x_1))/2$, in the $\Delta w = w - w_\mathrm{c}$ neighborhood of
the supercritical Hopf bifurcation of the coexistence fixed point
$\vect{u}^*_\mathrm{c}$. The dashed red curve ($\propto \sqrt{\Delta
w}$) is shown as a guide to the eye. (h) Period scaling in the
$\Delta w = w_\mathrm{d} - w$ neighborhood of the degenerate point
$w_\mathrm{d}$, with $T \propto \Delta w ^ {-\beta}$ where $\beta
\approx 0.8$.} \label{fig:N2cycles}
\end{figure}
%%%%%%%%%%%%%%%%%%%%%%%%%%%%%%%%%%%%%%%%%

%----------------------------------------
\subsection{Limit cycles and chaotic attractors}
\label{ap:dynamics_details}
%----------------------------------------

\paragraph{$N=2,3$ systems.}
For $1/T_x > N$, the four-dimensional $N=2$ system possesses four
fixed points: the $M=0$ origin $\fixp[0]{u}$, the coexistence $M=N$
fixed point, $\vect{u}^*_\mathrm{c} \equiv \fixp[{N}]{u} =
(1,1,1,1)$, and two dominance $M=1$ fixed points,
$\vect{u}^*_{\mathrm{d}_1} \equiv \fixp[{1}]{u_1} = (\fixpval{x}[1],
0, \fixpval{x}[1], 0)$ and $\vect{u}^*_{\mathrm{d}_2} \equiv
\fixp[{1}]{u_2} = (0, \fixpval{x}[1], 0, \fixpval{x}[1])$, respectively, with
$\fixpval{x}[1]=1/(1-w)$ (see Eqs.~(\ref{eq:M_winner})
and (\ref{eq:fixedpoint})).

\begin{figure} [t!]
 \centering
 \includegraphics[width=0.45\textwidth]{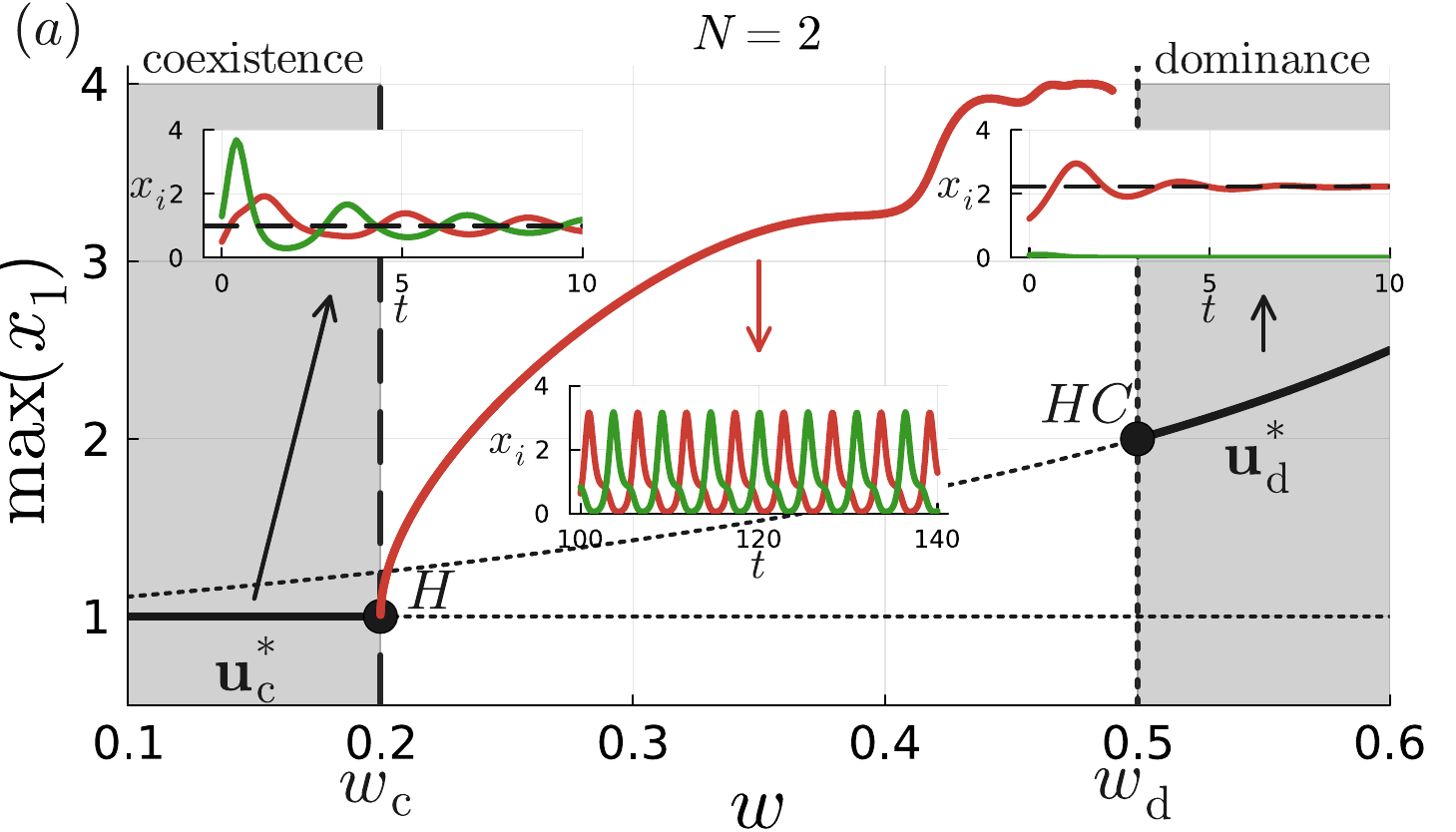}
 \includegraphics[width=0.45\textwidth]{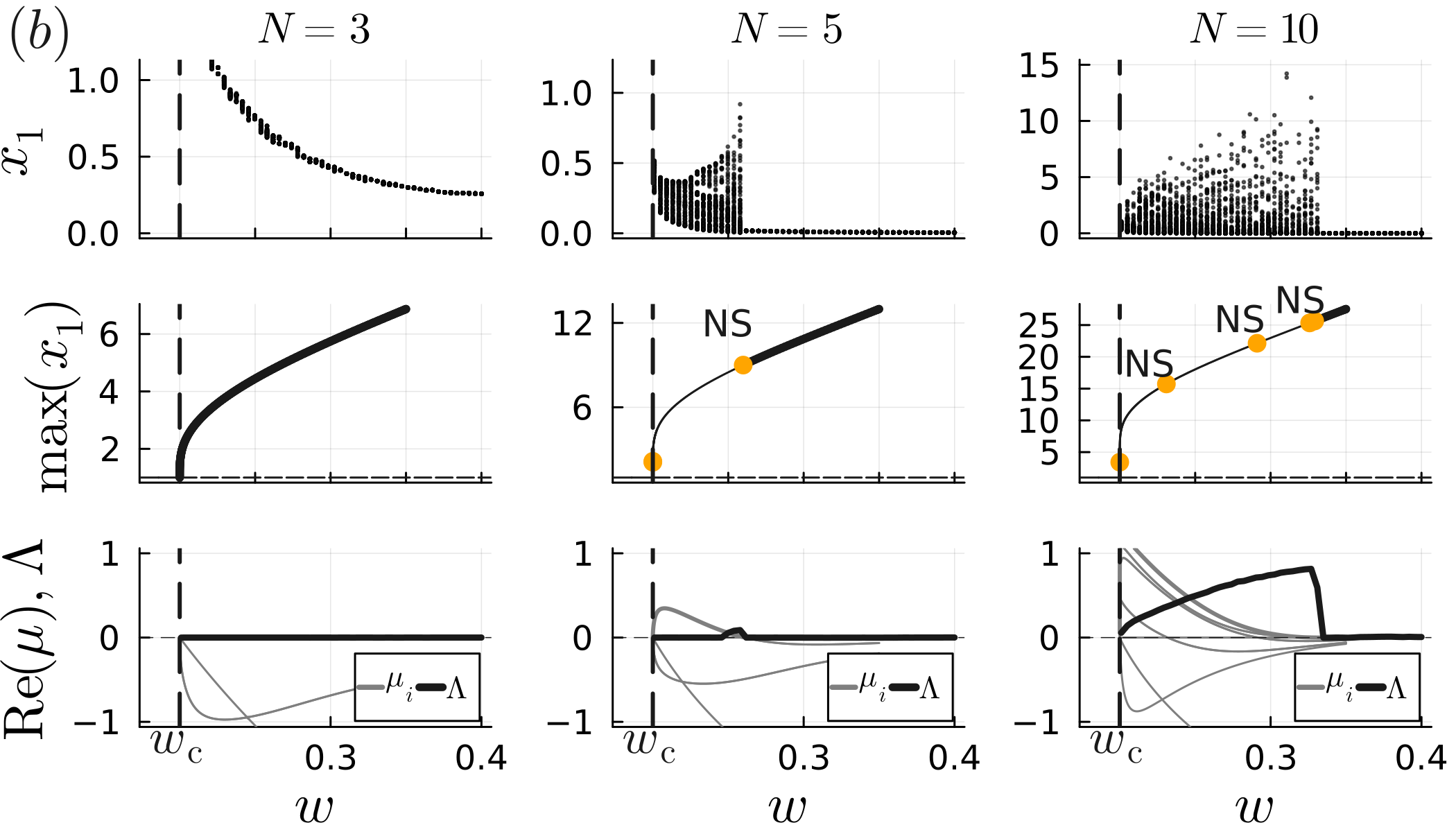}
 \caption{Bifurcation diagrams as a function of the control parameter~$w$ for different system sizes $N$ for normalized timescale $1/(T_x(N-1))=1.88$.
 (a) Limit cycles of the $N=2$ competitor system. Thick continuous/thin dotted curves denote stable/unstable fixed points, with their region of stability being indicated by the gray-shaded area (coexistence and dominance regimes). For the limit-cycle region, the amplitude of oscillations in the $x_1$ variable is indicated by the red curve. The insets show sample time series of the $x_1$ and $x_2$ activities approaching the coexistence/dominance fixed points in the left/right panels, and regular oscillations in the middle, respectively.
 (b) Bifurcations of limit cycles as a function of the relative coupling strength $w$ for different system sizes, $N=3,5,10$, showing only a single limit cycle out of $(N-1)!$. In each column, the orbit diagram of a limit cycle showing the $x_1$ components of the intersection points (run for $25000$ time units, discarding the first $20000$) with the Poincaré plane $x_2=1$ (top row), the bifurcation diagram obtained via numerical continuation (middle row), and the largest Lyapunov exponent $\Lambda$ together with the largest real part of the Floquet exponent $\Re(\mu)$ (bottom row) are shown, respectively. The NS points in the middle row mark the Neimark-Sacker bifurcation points along the limit cycle. Vertical dashed lines indicate the Hopf bifurcation point, $w_\mathrm{c} = 1/(T_x (N-1))$.
 }
 \label{fig:N2-10_bif} 
\end{figure}

From Eq.~(\ref{eq:eig}), it follows that the coexistence fixed point
($M = N$), with eigenvalues of type $\lambda_\mathrm{c}$ and
$\lambda_\mathrm{d}$, loses stability via a delay-induced
supercritical Hopf bifurcation at $w = w_\mathrm{c}$, as defined by
Eq.~(\ref{eq:coexistance_condition}). This transition gives rise to
limit-cycle oscillations in the WLC regime. These oscillations are
subsequently destroyed in a heteroclinic bifurcation at $w =
w_\mathrm{d}$, where the dominance fixed points become stable as the
limit cycle approaches and connects to the two symmetry-related
dominance states (compare the scaling properties near the
bifurcations in Fig.~\ref{fig:N2cycles} with the bifurcation diagram
obtained via numerical continuation in Fig.~\ref{fig:N2-10_bif}(a)).
For $N=3$ the behavior remains qualitatively
similar to the two-competitor case, except for the appearance of a
second, symmetry-related limit cycle (see Fig.~\ref{fig:cycles}(a)
and Fig.~\ref{fig:N2-10_bif}(b)). For $N=3$, two symmetry-related
cycles are created via degenerate Hopf bifurcations, similar to
those reported in \cite{sandor2015versatile}, differing only in the
order of sequential activation of the individual items.

\paragraph{$N=5,10$ competitors.} Increasing the number of
competitors further,  for $N=5$, from each of the $(N-1)! = 24$ limit cycles, tori emerge via
Neimark–Sacker bifurcations (indicated by positive Floquet exponents,
$\Re(\mu_i) > 0$), but no chaos is observed yet in the WLC region (see near-zero exponents in the $N=5$ system in panel (b) of Fig.~\ref{fig:N2-10_bif}).  With a further increase in the number of competitors (e.g., $N=10$), in addition to periodic dynamics with regular activity patterns (see
the white-colored region with $\Lambda = 0$ in
Fig.~\ref{fig:phase_diagN10}(b) and the corresponding time series in Fig.~\ref{fig:phase_diagN10}(a), panel C), irregular, chaotic
behavior also emerges, with positive Lyapunov exponents, $\Lambda > 0$
(compare the phase diagram in Fig.~\ref{fig:phase_diagN10}(b) with the
bifurcation diagrams in Fig.~\ref{fig:N2-10_bif}(b)).
The oscillations are chaotic over a wide range of relative timescales $T_x$ and coupling strengths
$w$ even after $T_\mathrm{tr}=10^4$ time units (see the red-colored
region with $\Lambda > 0$ in Fig.~\ref{fig:phase_diagN10}(b) and the
corresponding time series in Fig.~\ref{fig:phase_diagN10}(a), panel
B$_2$). Within this regime, both the duration and amplitude of
activations are irregular, and the order in which items receive
attention is also unpredictable. In the presence of noise, this
irregular dynamics remains the dominant behavior even in the
limit-cycle regime, as transient chaos in the $w \in (w_\mathrm{c},
w_\mathrm{d})$ interval can be rendered effectively permanent by
ubiquitous fluctuations in real systems. 

In the orbit diagrams shown in Fig.~\ref{fig:N2-10_bif}(b), the
variable $x_1$ is recorded at intersections with the Poincaré
section defined by $x_2 = 1$. The bifurcation diagrams based on
numerical continuation are obtained using the BifurcationKit.jl
Julia package~\cite{veltz2020}.

\paragraph{$N\gg 1$ systems.}
Note that, since the phase diagram introduced here has a
system-size--scaled parameter plane, most of its features are
system-size independent, and it looks almost the same (especially
for larger systems, compare the phase diagram for $N=10$ shown in
Fig.~\ref{fig:phase_diagN10}(b) with the one computed for $N=300$
items in Fig.~\ref{fig:phase_diagN300}(a)). Namely, the boundary
between the stable coexistence fixed-point regime and the WLC
dynamics is given by the same curve (thick dashed black curve), with
a shrinking WTA regime as $N$ is increased (the vertical dashed line
approaching the edge of the parameter plane,
$w_\mathrm{d}\rightarrow 1$ as $N\rightarrow\infty$ in
Eq.~(\ref{eq:WTA_condition})).  The increasing number of
symmetry-related limit cycles (see Appendix~\ref{ap:symmetries})
leads to exceedingly large transient lifetimes even within the
periodic regime of the $N=10$ system (see Fig.~\ref{fig:N10_lyap_ens}). Moreover, the chaotic region
expands further as the number of topics $N$ is increased (see
Sect.~1.2 of the Suppl. Mat.~\cite{supp} for $N=20,30,50,75$). For
$N=300$, the dynamics in the $w>w_\mathrm{c}$ regime after a
transient time of $T_\mathrm{tr}=500$ time units is practically
entirely chaotic (see Fig.~\ref{fig:phase_diagN300}(a)).
Therefore, irregular WLC constitutes the dominant dynamical regime in large systems with $N\gg 1$. These results suggest that delay-induced instability may represent a generic route to intermittent dynamics in high-dimensional competitive systems.

\subsection{Chaotic transients} 
\label{ap:chaotic_transients}
\begin{figure} [hbt!]
 \centering
 \includegraphics[width=0.9\textwidth]{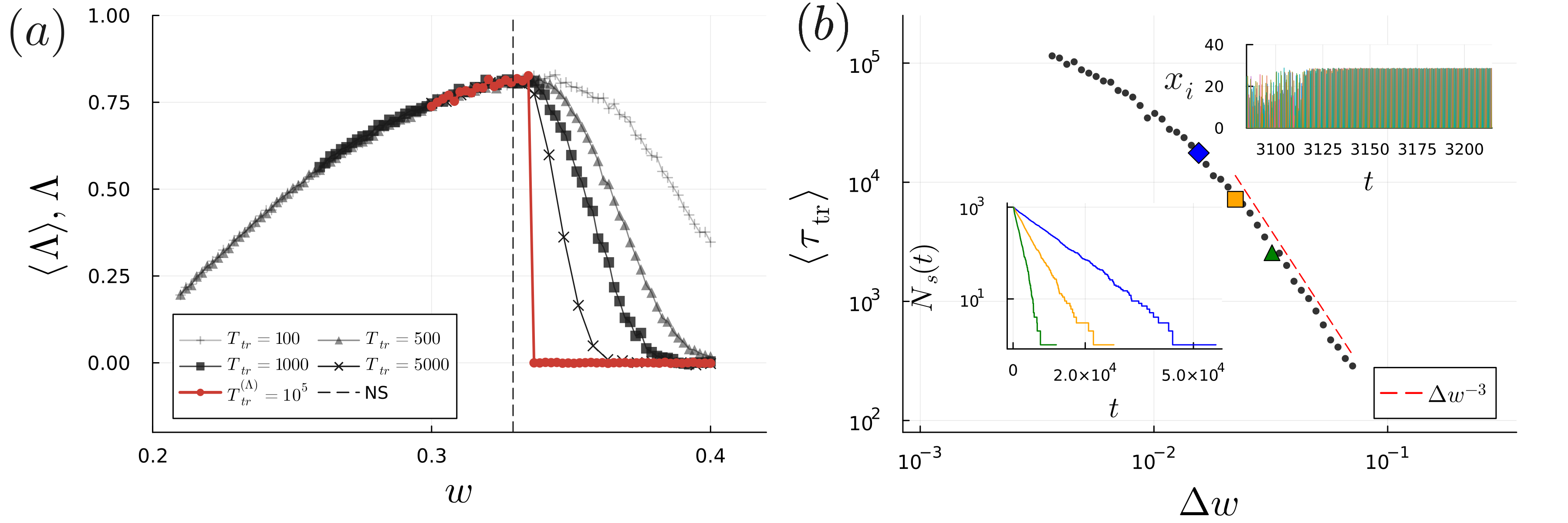}
  \caption{Transient chaos in the $N=10$ system for $y = 1/(T_x (N-1)) = 5$. 
  (a) Ensemble-averaged local Lyapunov exponents, $\langle \Lambda \rangle$ (defined in Eq.~(\ref{eq:ap_LLE})), for different transient times $T_\mathrm{tr}$ (grayscale curves with markers), compared with the largest Lyapunov exponent, $\Lambda$ (red-dotted curve), obtained via the Benettin method after a very long transient $T_\mathrm{tr}=10^5$, in the vicinity of the Neimark–Sacker bifurcation at $w_\mathrm{NS} \approx 0.329$ (dashed vertical line). The ensemble is initialized from states sampled on the chaotic attractor at $w = 0.3$. 
  (b) Average lifetime of transient chaos, $\langle \tau_\mathrm{tr} \rangle$, as a function of the distance from the bifurcation point, $\Delta w = w - w_\mathrm{NS}$. The red dashed line indicates a power law scaling, $\propto \Delta w^{-\beta}$, with $\beta = 3$, shown as a guide to the eye (note the log–log scale). The left inset presents the exponential decay of the number of survivors, $N_\mathrm{s}\propto e^{t/\langle\tau_\mathrm{tr}\rangle}$,  on the chaotic saddle as a function of time $t$ for three different parameter values, $w \approx 0.344,0.351,0.361$, indicated by the three markers in the main plot (green triangle, yellow square, and blue diamond), respectively. The right inset shows an example for $w \approx 0.361$ when the trajectory escapes and converges to a limit cycle.  } 
  \label{fig:N10_lyap_ens}
\end{figure}

We show here the properties of transient chaos for intermediate system sizes (our example is for $N=10$, see Fig.~\ref{fig:N10_lyap_ens}). For large coupling strengths $w$, chaos is suppressed due to the
stabilization of the $(N-1)!$ number of limit cycles, which sequentially visit
the still unstable dominance fixed points, with different visitation
orders for each cycle.  In this domain, transient chaos precedes
limit cycle oscillations. This is demonstrated below by two
different methods. 
Firstly, to quantify the persistence of chaotic dynamics following
the destabilization of the attractor at $w = w_\mathrm{NS} \approx
0.329$ for $N=10$, we employ an ensemble-based method to calculate
the average local Lyapunov Exponent (LLE), $\langle\Lambda\rangle$.
The procedure is defined as follows~\cite{tel2006chaotic}:

\begin{enumerate}
    \item Ensemble initialization: a set of $N_\mathrm{ens}=10^3$
initial conditions are distributed according to the natural measure
of the former chaotic attractor for $w=0.3<w_\mathrm{NS}$.  \item
Transient evolution: setting a new parameter $w$, each trajectory is
allowed to evolve for a varying transient time $T_\mathrm{tr}$. No
trajectories are removed from the ensemble, even if they have
settled into a stable limit cycle.
    \item Local expansion measurement: at time $T_\mathrm{tr}$, an
infinitesimal perturbation $\delta
\mathbf{u}_i(T_\mathrm{tr})=\delta_0=10^{-9}$ is applied to each
reference trajectory $\mathbf{u}_i(t)$ with
$i=1,\dots,N_\mathrm{ens}$. Then the logarithmic divergence of the
absolute value of the perturbation over a short interval $\Delta
t=10$ is computed as: \begin{equation}
        \langle \Lambda(T_\mathrm{tr}) \rangle = \frac{1}{N_\mathrm{ens}} \sum_{i=1}^{N_\mathrm{ens}} \left[ \frac{1}{\Delta t} \ln \frac{|\delta \mathbf{u}_i(T_\mathrm{tr} + \Delta t)|}{\delta_0} \right]
        \label{eq:ap_LLE}
    \end{equation}
    \item The above procedure is repeated for control parameters $w\in(0.2,0.4)$.
    This yields a time-dependent average LLE that captures the
residual chaos in the system (see Fig.~\ref{fig:N10_lyap_ens}(a)).
\end{enumerate}

In the chaotic attractor regime, the ensemble-averaged LLEs obtained
for different, but relatively short transient times, $T_\mathrm{tr}
\leq 5000$, are consistent with each other and also agree with the
largest Lyapunov exponent, $\Lambda$, computed using the Benettin
method for a single long trajectory after a much longer transient,
$T_\mathrm{tr} = 10^5$.

Furthermore, in the limit-cycle regime, ensemble-averaged LLEs,
$\langle \Lambda(T_\mathrm{tr}) \rangle$, computed with relatively
short transients, $T_\mathrm{tr} = 100, 500$, follow the same
envelope as a function of the coupling strength $w$, even beyond the
bifurcation point, indicating the presence of transient chaos.
However, as the transient time increases, these estimates eventually
converge to the Benettin-method value, $\Lambda = 0$ (see
Fig.~\ref{fig:N10_lyap_ens}(a)).  The destabilization of the
attractor coincides with a Neimark-Sacker (NS) bifurcation of the
target limit cycle, $w_\mathrm{NS}\approx 0.329$ (up to the
parameter resolution used for numerical continuation). 
The local expansion measurement given by Eq.~(\ref{eq:ap_LLE}) 
can also be approximated by running the Benettin algorithm 
for a short time interval ($\Delta t = 10$). 
Here, we used functions from the ChaosTools.jl library within DynamicalSystems.jl \cite{DynamicalSystems}.

Secondly, by computing the average lifetime of transient chaos, $\langle
\tau_\mathrm{tr} \rangle$, from the escape rate of trajectories from
the chaotic saddle to one of the stable limit
cycles~\cite{tel2008,Sandor2013}, and analyzing its dependence on
the distance from the bifurcation point, $\Delta w = w -
w_\mathrm{NS}$, a power law scaling is observed, $\langle
\tau_\mathrm{tr} \rangle \propto \Delta w^{-\beta}$, with $\beta
\approx 3$ and a divergence as $\Delta w \to 0$ (see
Fig.~\ref{fig:N10_lyap_ens}(b)).

Our results demonstrate two distinct regimes:

\begin{itemize} \item Short Transients, $T_\mathrm{tr} < \langle
\tau_\mathrm{tr} \rangle$: The calculated $\langle
\Lambda(T_\mathrm{tr}) \rangle$ remains positive and nearly
identical to the Lyapunov exponent of the pre-crisis attractor (see
e.g. $T_\mathrm{tr}=100$ in Fig.~\ref{fig:N10_lyap_ens}(a)). This
confirms the existence of a chaotic saddle—a non-attracting
invariant set that retains the stretching-and-folding dynamics of
the original attractor.
    \item Long Transients, $T_\mathrm{tr} > \langle \tau_\mathrm{tr} \rangle$: The average $\langle \Lambda(T_\mathrm{tr}) \rangle$ asymptotically approaches zero (see e.g. $T_\mathrm{tr}=5000$ in Fig.~\ref{fig:N10_lyap_ens}(a)). This represents the leakage of the ensemble from the saddle into the stable limit cycles. The transition of the limit cycle via an NS bifurcation acts as the global mechanism for this escape.
\end{itemize}

\begin{figure} [hbt!]
 \centering
 \includegraphics[width=0.7\textwidth]{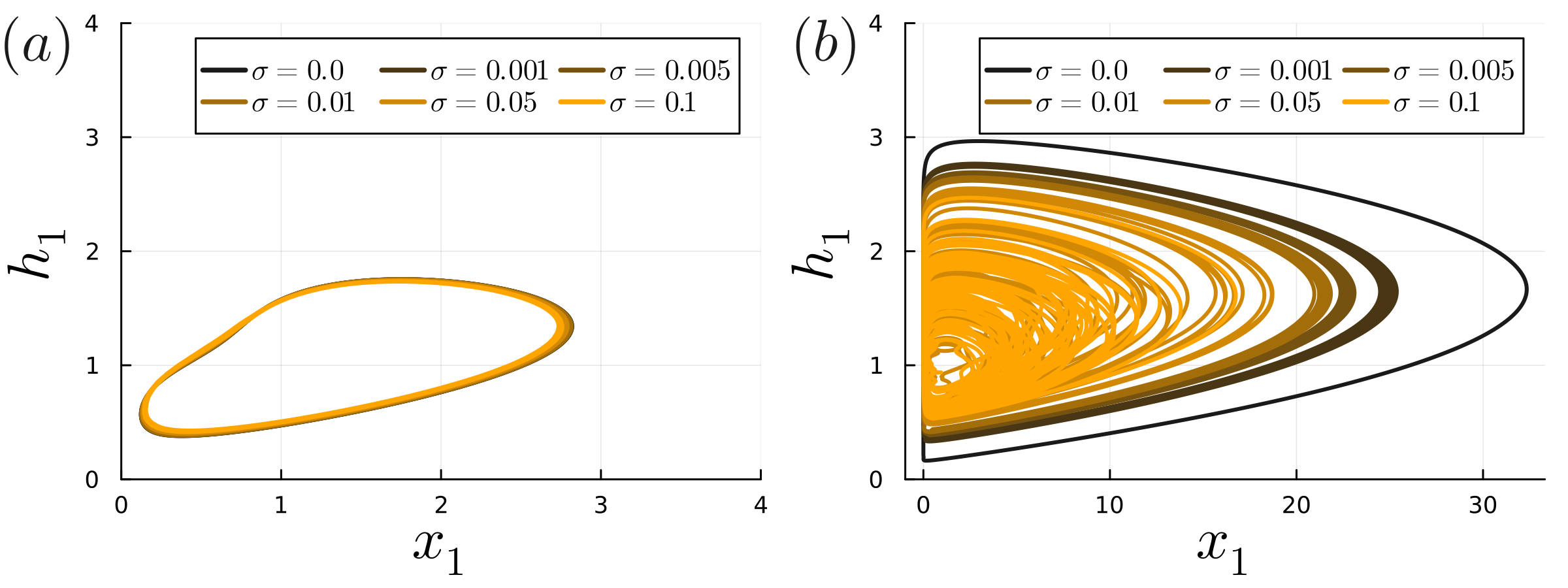}
 \caption{Trajectories with additive noise projected to the $(x_1,h_1)$ plane for $1/T_x(N-1)=5$ and coupling strength $w$ within the stable limit cycle regime. The added noise is uniform with amplitude $\sigma$ (added to the right-hand side of the equations for $x_i$ in Eq.~(\ref{eq:rescaled})).  (a) For $N=2$ competitors and $w=0.3$, trajectories stay near the limit cycle, regardless of the amplitude of the noise. (b) In the $N=10$ case with $w=0.4$, trajectories with a noise amplitude larger than a critical value, $\sigma\geq 0.05$, are knocked around on the chaotic saddle. } 
 \label{fig:noise}
\end{figure}

\paragraph{Noise-induced chaos.}
To demonstrate that chaotic behavior is the generic dynamics in larger delayed Lotka-Volterra systems, we examine a more realistic scenario in which the dynamics is perturbed by noise. Specifically, white noise with uniform distribution and amplitude $\sigma$ is added to the right-hand side of the equations for $x_i$ in Eq.~(\ref{eq:rescaled}).
As expected, in small systems, even relatively large noise amplitudes lead only to slight deviations from the limit-cycle attractor (see Fig.~\ref{fig:noise}(a)). In contrast, for larger systems exhibiting transient chaos, there exists a critical noise level above which the chaotic dynamics becomes effectively permanent (similarly to the cases reported in~\cite{tel2008,Sandor2013}), resulting in persistent irregular oscillations in the attention dynamics (see Fig.~\ref{fig:noise}(a)).

\subsection{Numerical integration of large systems}
\label{ap:numerics}
\paragraph{Alternative form of the equations using a logarithmic transform.}
To avoid floating-point underflow in large systems with $N=300$ topics, where trajectories in the chaotic WLC regime may approach zero, $x_i\rightarrow 0$ and $L_i\rightarrow 0$, we solve an alternative form of Eq.~(\ref{eq:delay_split_orig}) obtained through the transformation $l_i=\ln L_i$, 
\begin{equation}
\begin{aligned}
    \dot{l}_i &= \frac{1}{L_i} \dot{L}_i = r_p \left( 1 - \frac{r_c}{K}Y_i - c \sum_{j\neq i}^N L_j \right)\,,  \\
    \dot{Y}_i &= e^{l_i} - \alpha Y_i\,.
\end{aligned}
\label{eq:logL}
\end{equation}
Without this transformation, inactive items may fall below machine precision, underflow to zero, and never recover.

Since the largest Lyapunov exponent is invariant under smooth transformations~\cite{tel2006chaotic}, the logarithmic formulation can be used for its numerical computation without affecting the resulting value. Note, however, that the largest Lyapunov exponents of the $(x,h)$ and $(L,Y)$ systems are related by $\Lambda=\Lambda_L / \alpha$, where the subscript $L$ denotes the Lyapunov exponent of the $(L,Y)$ system (see Sect.~1.2 in the Suppl. Mat.~\cite{supp}).
The formulation (\ref{eq:logL}) was used to produce Fig.~\ref{fig:phase_diagN300}. In this work, we used the DynamicalSystems.jl library to obtain trajectories and Lyapunov exponents \cite{DynamicalSystems}. We used the Tsitouras $5/4$ solver to integrate the trajectories (default in the library) \cite{Tsitouras2011}.

\section{Lifetime distributions and increasing cultural yield} \label{ap:lifetimes}

\begin{figure} [hbt!]
    \centering
    \includegraphics[width=0.9\linewidth]{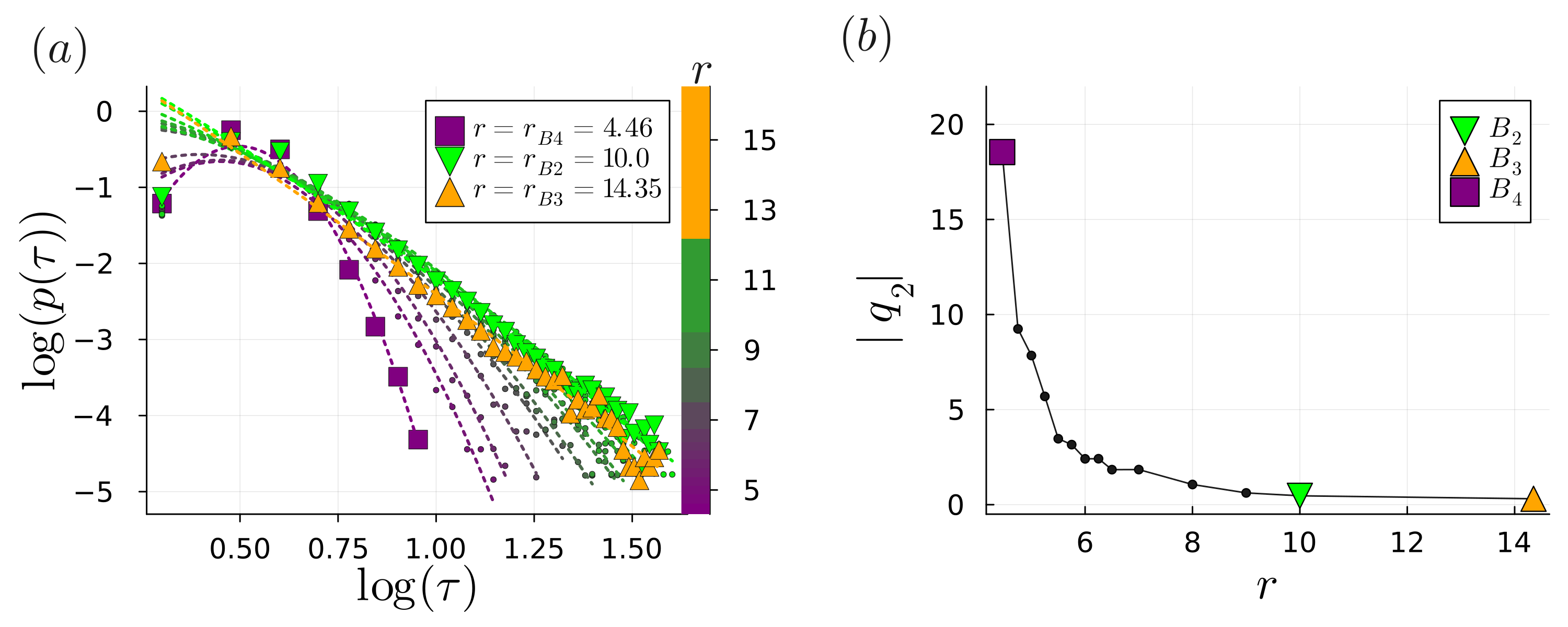}
    \caption{Shift of lifetime distributions as production/consumption rates $r$ are increased in system (\ref{eq:delay_split_orig}) with $N=300$ items. (a) Lifetime distributions from charts with charting time $\Delta t_L = 2.0$, for a range of $r$ values on the cultural yield curve (see white dotted curve $1/(T_x(N-1))\propto 1/w - 1$ inside the chaotic region on Fig.~\ref{fig:phase_diagN300} where $r$ increases in the direction along the $B_4$, $B_2$, and $B_3$ points). The distributions corresponding to these three selected values are highlighted with the same markers and colors as seen on Fig.~\ref{fig:phase_diagN300} in the main text. The fitted parabolas are shown by the thin dashed curves with color-coded $r$ values (see the colorbar).
    (b) Quadratic coefficient $|q_2|$ in terms of the cultural yield $r$. With increasing production/consumption rate $r$, the lifetime distributions shift from variance-sensitive (log-normal) to mean-dominated (power law) regimes. }
    \label{fig:lifetimes_and_fitparams}
\end{figure}

The transition between the log-normal and power-law-like distributions,
\begin{equation}
\label{eq:p_log_normal}
p(\tau) \sim \frac{1}{\tau}\,\exp(q_0 + q_1\log\tau + q_2(\log\tau)^2)\,, 
\qquad \text{and}\quad 
p(\tau) \sim \tau^{q_1-1}\,,
\end{equation}
respectively, can be quantified by computing the absolute value of the quadratic coefficient, $|q_2|$, obtained by fitting the parabola
\begin{equation}
\log(p(\tau)) = q_0 + (q_1-1)\log\tau + q_2(\log\tau)^2 + \mathrm{const.}
\end{equation}
to the $\log p(\tau)$ versus $\log\tau$ data for different values of the cultural yield $r$ (see Fig.~\ref{fig:lifetimes_and_fitparams}(a)). The fitted coefficient $|q_2|$ decreases abruptly for $r<6$, converging eventually to 0 for larger production rates $r$ (Fig.~\ref{fig:lifetimes_and_fitparams}(b)).
In order to capture the transition to a power-law tail, we used a fitting algorithm with weighted least squares \cite{Strutz2011,Geman92}. The corresponding loss function for $N_\mathrm{dat}$ data points is given by
\begin{equation}    
\mathcal{L} = \sum_{i=1}^{N_\mathrm{dat}} a_i R_i , \quad R_i = (D_i - f(x_i))^2,
\end{equation}
where residuals $R_i$ are mapped to weights as
\begin{equation}
    \quad a_i = \frac{1}{R_i + \bar{R}} , \quad \bar{R} = \frac{1}{N_\mathrm{dat}} \sum_{j=1}^{N_\mathrm{dat}} R_j,
\end{equation}
in a way that outliers (data points from the distribution with large deviation from the quadratic model) get smaller weights in the weighted sum of residuals. The mean of squared residuals $\bar{R}$ acts as a lower bound, so residuals close to zero do not lead to disproportionally large weights. The weights are normalized and adjusted iteratively until convergence, using the iteratively reweighted least squares (IRLS) method.
\end{document}

\maketitle

\begin{abstract}	

\end{abstract}
%%%%%%%%%%%%%%%%%%%%%%%%%%%%%%%%%%%%%%%%%%%%%%%%%%%%%

%%%%%%%%%%%%%%%%%%%%%%
\section{Stability analysis}
%%%%%%%%%%%%%%%%%%%%%%

\subsection{Classical Lotka-Volterra system with symmetric coupling} 
\label{sec:ap_classical}

\begin{figure} [hbt!] 
    \centering
    \includegraphics[width=0.9\textwidth]{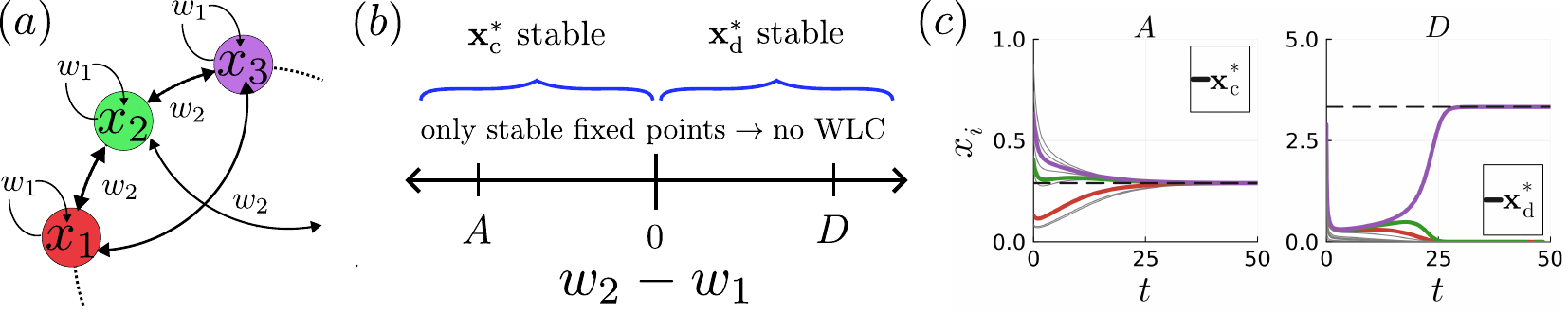}
    \caption{Dynamics of the symmetric Lotka-Volterra system without delay, denoting with $w_1$ and $w_2$ the strength of self-interaction and of the competition, respectively.
    (a) Sketch of the interactions and coupling parameters between items $x_i$ for $i=1,2,3$.
    (b) Schematic presentation of the stable parameter regimes of the coexistence $\vect{x}^\ast_\mathrm{c}$ and dominance $\vect{x}^\ast_\mathrm{d}$ fixed points, in terms of the relative competition strength $w_2-w_1$. No winnerless competition is possible. 
    (c) Time series of the system with $N=10$ competitors. For simplicity, only the first three activities $x_i$ with $i=1,2,3$ are shown with the corresponding colors as in panel (a), while the others are ploted in gray. Low competition with $w_2 < w_1$ results in coexistence, with stable fixed point $\vect{x}^\ast_\mathrm{c}$ (panel A, $w_1=0.75$ and $w_2=0.3$).  Higher competition with $w_1 < w_2$ leads to a winner-takes-all dynamics, characterized by the 10 symmetry-related dominance fixed points, $\vect{x}^\ast_\mathrm{d}$ (panel D, $w_1=0.3$ and $w_2=0.75$). }
    \label{fig:nodelay_LV}
\end{figure}

We identify delay as a minimal ingredient for generating WLC in symmetric systems. To illustrate that delay can induce WLC in a symmetric LV system, and that it is absent without delay, we examine the possible dynamics of a classical symmetric LV system without delay terms. In this setting, identical competitor equations are coupled symmetrically and uniformly, with instantaneous self-interaction rather than delayed feedback:
\begin{equation} \label{eq:nondelay}
    \dot{x}_i = rx_i \left( 1 - w_1x_i - w_2\sum_{j \neq i}^N x_j \right)\,,
\end{equation}
where $x_i$ denotes the population of the $i$-th species, with $i=1,\dots,N$, while $w_1$ and $w_2$ denote the strengths of self-interaction and competition, respectively (for a schematic illustration of the system, see Fig.~\ref{fig:nodelay_LV}(a)). Note that the above parameter setting, which distinguishes between the strengths of self-interaction $w_1$ and competition $w_2$, represents a slightly more general version of the delayed LV system given by Eq.~(1) in the main paper. The latter is recovered by setting $w_1= 1-w$ and $w_2=w$.

The dynamics of \eqref{eq:nondelay} is well-known: since periodic orbits can be ruled out, asymptotic behaviour is restricted to stable fixed points. The phase point can only converge to the coexistence fixed point, or one of the $N$ single-winner fixed points \cite{goh1977global,Varona2002,Cohen1983}. 

\paragraph{Fixed points.}
The fixed points of the system are 
\begin{equation}
\fixp[M]{x} = (\fixpval{x}[M],\fixpval{x}[M],\dots,\fixpval{x}[M],0,\dots,0)\,,\qquad \fixpval{x}[M] = 1/(w_1+(M-1)w_2)\,,
\end{equation}
or any permutation thereof containing $M$ number of nonzero values. The nonzero values are given as the solutions of the homogeneous linear system,
\begin{equation} \label{eq:ap_fp_sys_classic}
\begin{aligned}
  w_1 \cdot \fixpval{x}[M][1] + w_2(\fixpval{x}[M][2]+\fixpval{x}[M][3]+\fixpval{x}[M][4]+\dots+\fixpval{x}[M][M]) &= 1 \\
  w_1 \cdot \fixpval{x}[M][2] + w_2(\fixpval{x}[M][1]+\fixpval{x}[M][3]+\fixpval{x}[M][4]+\dots+\fixpval{x}[M][M]) &= 1\\
    &\vdots \\
  w_1 \cdot \fixpval{x}[M][M] + w_2(\fixpval{x}[M][1]+\fixpval{x}[M][2]+\fixpval{x}[M][3]+\dots+\fixpval{x}[M][{M-1}]) &= 1\,,
\end{aligned}
\end{equation}
as $\fixpval{x}[M][1] = \fixpval{x}[M][2] =...= \fixpval{x}[M][M] \equiv \fixpval{x}[M]$. This becomes obvious if one takes the difference of any two equations from \eqref{eq:ap_fp_sys_classic}, resulting in $(w_1 - w_2)(\fixpval{x}[M][i] - \fixpval{x}[M][j]) = 0$.
Since the $w_1=w_2$ case leads to an infinite number of solutions, the above can only be satisfied if $\fixpval{x}[M][i]=\fixpval{x}[M][j]$. Therefore, the nonzero values need to be identical.
Thus, the number of fixed points within a group is given by the number of $M$-combinations of the $i$ indices out of $N$, denoted by the binomial coefficient $\binom{N}{M}$.

\paragraph{Linear stability}

We now turn to building and calculating the eigenvalues of the Jacobian at the fixed points.
After calculating the partial derivatives 
\begin{equation}    
     \frac{\partial{\dot{x_i}}}{\partial{x_i}} = r \left( 1 - 2w_1x_i - w_2\sum_{j \neq i} x_j  \right) \equiv A_i, \qquad \frac{\partial{\dot{x_i}}}{\partial{x_j}}\Bigr|_{\substack{i\neq j}} = -rw_2x_i \equiv B_i,
\end{equation}
one can plug in the ordered $\fixp[{M}]{x} = (\fixpval{x}[M],\fixpval{x}[M],...,\fixpval{x}[M],0,...,0)$ without any loss of generality. Depending on $x_i$, the matrix elements can be
\begin{equation}
    A_i = \begin{cases}
        A_x \equiv r(1 - 2w_1 \cdot \fixpval{x}[M] - w_2(M-1)\cdot \fixpval{x}[M])\,, \quad \text{for} \quad i=1,\dots,N-M \,,\\
        A_0 \equiv r(1 - w_2M \cdot \fixpval{x}[M] )\,, \quad \text{for} \quad i=M+1,\dots,N\,,\\
    \end{cases}
\end{equation}
\begin{equation}
    B_i = \begin{cases}
        B \equiv -rw_2 \cdot \fixpval{x}[M]\,, \quad \text{for} \quad i=1,\dots,N-M\,, \\
        0\,, \quad \text{for} \quad i=M+1,\dots,N\,.\\ 
    \end{cases}
\end{equation}
The Jacobian takes a block matrix form:
\begin{equation}
\textbf{J}=
 \begin{pmatrix}
    A_x & B & \dots & B & B & B & \dots & B\\
    B & A_x & \dots & B & B &\ddots & & \vdots\\
    \vdots & & \ddots & \vdots \\
    B &    & \dots     & A_x & B & & \dots & B \\
    \hline 
    0 & 0 & \dots & 0 & A_0 & 0 & \dots & 0 \\
    0 & \ddots  &  & 0 & 0 & A_0 & \dots & 0\\
    \vdots &  & & \vdots & \vdots &  & \ddots & \vdots \\ 
    0 & 0 &\dots & 0 & 0 & & \dots &A_0 
    \end{pmatrix} \equiv
    \left(
    \arraycolsep=5.4pt\def\arraystretch{2}
    \begin{array}{c|c}
    \vect{A_B} & \vect{B}  \\
    \hline
    \vect{O}  & \vect{A_0}
    \end{array}
    \right)
    \end{equation}
Eigenvalues can be calculated from the characteristic polynomial:
\begin{equation} \label{eq:ap_charpoly_classic1}
    \det(\vect{J - \lambda I_N}) = 0 
    \qquad\Rightarrow\qquad
    \det \left(
    \arraycolsep=5.4pt\def\arraystretch{2}
    \begin{array}{c|c}
    \vect{A_B - \lambda I_M} & \vect{B}  \\
    \hline
    \vect{O}  & \vect{A_0 - \lambda I_{N-M}}
    \end{array}
    \right) = 0\,.
\end{equation}
% Using the identity $\det \big(\begin{smallmatrix}
%   \vect{A} & \vect{B}\\
%   \vect{C} & \vect{D}
% \end{smallmatrix}\big) = \det(\vect{D})\cdot\det(\vect{A} - \vect{B} \vect{D}^{-1} \vect{C})$ 
The determinant in (\ref{eq:ap_charpoly_classic1}) has a block upper triangular structure, hence it can be reduced  via Schur complements to a product of determinants \cite{Abadir_Magnus_2005}, $\det(\vect{A_0 - \lambda I_{N-M}}) \cdot \det(\vect{A_B - \lambda I_M}) = 0 $. The second determinant can be calculated for example using an identity for circular matrices. In summary, the characteristic polynomial can be factorized as:
\begin{equation}
        \underbrace{(A_0-\lambda)^{N-M}}_{\lambda_a} ~ \cdot ~ \biggl[ \underbrace{(A_x-\lambda)+B(M-1)}_{\lambda_b} \biggr] ~ \cdot ~ \underbrace{(A_x-\lambda - B)^{M-1}}_{\lambda_c} = 0\,,
\end{equation}
leading to three kinds of eigenvalues:
\begin{equation}  \label{eq:ap_classic_eig}
\begin{aligned}
\lambda_a &= A_0 =  \frac{r(w_1-w_2)}{w_1+(M-1)w_2}, \qquad m(\lambda_a) = N-M\,, \\
\lambda_b &= A_x + B(M-1) = -r , \qquad m(\lambda_b) = 1\,, \\
\lambda_c &= A_x - B = \frac{r(w_2-w_1)}{w_1+(M-1)w_2},  \qquad m(\lambda_c) = M-1\,,
\end{aligned}
\end{equation}
where $m(\lambda)$ denotes the algebraic multiplicity (see Table \ref{tb:classic_eigspectra}). From \eqref{eq:ap_classic_eig} we can conclude the following:

\begin{enumerate}[label=(\roman*)]
    \item The coexistence fixed point $\vect{x}^*_{\mathrm{c}} \equiv \fixp[{N}]{x}$ is stable if $w_1 > w_2$ since it's spectrum only contains eigenvalues of type $\lambda_b$ and $\lambda_c$.
    \item The single-winner fixed points $\vect{x}^*_{\mathrm{d}} \equiv \fixp[{1}]{x}$ is stable if $w_1 < w_2$ since it's spectrum only contains eigenvalues of type $\lambda_a$ and $\lambda_b$.
    \item The M-winner fixed points, $\fixp[{M}]{x}$, where $M=2,3,...,N-1$, cannot be stable since their spectrum contains all eigenvalue types and the stability conditions for $\lambda_a$ and $\lambda_c$ contradict each other.
\end{enumerate}

\begin{table}[htbp!] 
\centering 
\def\arraystretch{1.5}%
\begin{tabular}{c|c|c|c}
    \hline
    $M$ & $\lambda_a$ & $\lambda_b$ & $\lambda_c$  \\
    \hline
     $0$ & $N$ & $\cross$ & $\cross$   \\
     \hline
     $1$ & $N-1$ & $1$ & $\cross$   \\ 
     \hline
     $2,3,...,N-1$ & $N - M$ & $1$ & $M-1$   \\
     \hline
     $N$ & $\cross$ & $1$ & $N-1$   \\
     \hline
\end{tabular}
\caption{Eigenvalue spectrum of fixed point groups $\fixp[{M}]{x}$. When the eigenvalue type is present in the spectrum for a particular $M$, the multiplicity is shown, while missing eigenvalue types are denoted by the $\cross$ symbols. Also note that each row adds up to $N$ eigenvalues in total.}
\label{tb:classic_eigspectra}
\end{table}

In summary, no WLC is present in this kind of system. Possible dynamics are converging to coexistence or single-winner fixed points, depending on the relation of the self-interaction strength and the coupling strength with other competitors (see panels (b) and (c) of Fig.~\ref{fig:nodelay_LV}).

%%%%%%%%%%%%%%%%%%%%%%%%%%%%%%%%%%
\subsection{Delayed Lotka-Volterra equations}
\label{sec:ap_fixpoints}

Our basic model is
\begin{equation}
    \begin{aligned} \label{eq:rescaled}
       T_x\,\dot{x}_i &= x_i \big[ 1 - (1 - w) h_i - w \langle x \rangle_i \big]\,, 
        \qquad\quad \langle x \rangle_i &= \frac{1}{N-1}\sum_{j\neq i} x_j\,, \\
    T_h\,\dot{h}_i &=  x_i -  h_i ,
    \end{aligned}
\end{equation}
where  $h_i$ is the memory of item $i$. 
The parameter $w\in[0,1]$ regulates the relative weight of
self- and mutual interaction, with $T_x, T_h>0$ being the
respective time scales. Without loss of generality, in the following, we fix $T_h=1$.

\paragraph{Fixed points.} Fixed points $\vect{u}^{*} = (\vect{x}^{*}, \vect{h}^{*} ) =(x_1^{*},x_2^{*},...,x_N^{*},h_1^{*},h_2^{*},...,h_N^{*})^\top$ of the system (\ref{eq:rescaled}) satisfy
\begin{align}
 \dot{{x}}_i &= 0 \qquad\Rightarrow\qquad x^{*}_i \left[ 1 - (1 - w) h^{*}_i - w \langle{x}^{*}\rangle_i \right] = 0\,,   \label{eq:fpsystem1} \\
\dot{{h}}_i &= 0  \qquad\Rightarrow\qquad  x^{*}_i -  h^{*}_i = 0,  \label{eq:fpsystem2}
\end{align}
for every $ 1 \leq i  \leq N$. Finding the fixed points is actually the same problem as it would be for a set of classical Lotka-Volterra equations: the fixed points $\fixp[{M}]{u}$ can be classified into $N+1$ groups, each corresponding to $M = 0,1,...,N$ competitors having nonzero activity levels, while the other $N-M$ being set to zero. The case $M = 0$ corresponds to the origin (total extinction), while $M = N$ would mean coexistence of all competitors, where all activities are constant and identical. With $M = 1$ only one activity level stays nonzero, the others vanish, thus "dominating" the others (this is the so-called "winner-takes-all" situation).\\
Equally-weighted coupling greatly simplifies finding the fixed points of the system, namely, it reduces it to finding the vertices of an $N$-dimensional cube in the $x$-subspace. This becomes clear as Eq.~\eqref{eq:fpsystem1} can be satisfied in two ways: either $\fixpval{x}[M][i] = 0$ or the entire expression inside the square brackets in (\ref{eq:fpsystem1}) vanishes.
% \begin{enumerate}
%     \item $x_i^* = 0$ \\
%     \item $1 - (1 - w) y_i^* - w \bar{x}_i^*  = 0$. \label{eq:parantheses}
% \end{enumerate}
 Obtaining the nontrivial value at the fixed point requires solving the linear system of $M$ equations. Using condition \eqref{eq:fpsystem2}, we get
\begin{equation} \label{eq:ap_fplinsyseq1}
\begin{aligned}
  (1-w)\cdot \fixpval{x}[M][1] + \frac{w}{N-1}(\fixpval{x}[M][2] + \fixpval{x}[M][3] +...+ \fixpval{x}[M][M]) &= 1 \\
  (1-w) \cdot \fixpval{x}[M][2] + \frac{w}{N-1}(\fixpval{x}[M][1] +  \fixpval{x}[M][3] +...+ \fixpval{x}[M][M]) &= 1\\
    &\vdots \\
  (1-w) \cdot \fixpval{x}[M][M] + \frac{w}{N-1}(\fixpval{x}[M][1] + \fixpval{x}[M][2] +...+ \fixpval{x}[M][M-1]) &= 1,
\end{aligned}
\end{equation}
or in matrix form
\begin{equation} \label{eq:ap_fplinsyseq2}
    \begin{pmatrix} 
     1-w & \frac{w}{N-1} & \dots & \frac{w}{N-1} \\
     \frac{w}{N-1} & 1-w & \dots & \frac{w}{N-1} \\
    \vdots & & \ddots & \vdots \\
     \frac{w}{N-1} &    & \dots     & 1-w 
    \end{pmatrix} 
     \begin{pmatrix}
     \fixpval{x}[M][1] \\
     \fixpval{x}[M][2] \\
     \vdots \\
     \fixpval{x}[M][M]
    \end{pmatrix} = 
     \begin{pmatrix}
     1 \\
     1 \\
     \vdots \\
     1
    \end{pmatrix}\,.
\end{equation}
Eq.~\eqref{eq:ap_fplinsyseq2} has a unique solution only if $1-w \neq \frac{w}{N-1}$ such that $i=1,\dots,M$ competitors have $\fixpval{x}[M][i]=\fixpval{x}[M]>0$ activities, while for the rest $i=M+1,\dots,N$ competitors is $\fixpval{x}[M][i]=0$,
\begin{equation} 
    \label{eq:ap_M_winner}
    \fixp[{M}]{x} = ( \underbrace{\fixpval{x}[M],\fixpval{x}[M],...,\fixpval{x}[M]}_M,\underbrace{0,0,...,0}_{N-M} )\,,
\end{equation}
This can be justified by subtracting the $i$-th equation from the $j$-th equation in system \eqref{eq:ap_fplinsyseq1}, similarly to the non-delay version in Sec.~\ref{sec:ap_classical}. 
This means that all nonzero competitors have the same activity levels and memory values $\fixpval{x}[M][i] = \fixpval{h}[M][i] = \fixpval{x}[M]$ in equilibrium, given as:
\begin{equation} 
\fixpval{x}[M] = \begin{cases}    
\label{eq:ap_fixedpoint}
    \frac{N - 1}{N-1 + wM - wN}, \quad \text{if} \quad M \neq 0 ~ \text{and} ~ N > 1 \\[10pt]
    \frac{1}{1-w}, \quad \text{if} \quad  N=1 .
\end{cases} 
\end{equation}
The fixed points where $M$ number of activity levels are nonzero, can be denoted by $\fixp[{M}]{u}=(\fixp[{M}]{x},\fixp[{M}]{h})$. Taking all combinations into consideration, there are in total $\sum_{M=0}^N \binom{N}{M} = 2^N$ fixed points. 

\paragraph{Linear stability.} Since every fixed point with the same $M$ must have the same stability conditions due to symmetry, we can talk about stability of entire groups of fixed points for a given $M$. 
For the linear stability of the system Eq.~(\ref{eq:rescaled}), one can obtain the eigenvalues of the Jacobian evaluated at the fixed points in the general $N$-competitor case. When computing the Jacobian $\mathbf{J}$, one can group the equations for $x_i$ and $h_i$ variables, respectively, and write it in a block matrix form:
\begin{equation}
\mathbf{J}=
\left(
\arraycolsep=2.4pt
\begin{array}{c|c}
\partial{\dot{x_i}} / \partial{x_j} & \partial{\dot{x_i}} / \partial{h_j}  \\
\hline
\partial{\dot{h_i}} / \partial{x_j} & \partial{\dot{h_i}} / \partial{h_j}  \\
\end{array}
\right)\,,
\end{equation}
where each block matrix is $N \times N$ dimensional.
After calculating the partial derivatives
\begin{align*}
    \frac{\partial{\dot{x}_i}}{\partial{x_j}}\Bigr|_{\substack{i\neq j}} &= \frac{-gw}{N-1}x_i\,, \quad \frac{\partial{\dot{x}_i}}{\partial{x_i}} = g[1-(1-w)h_i-w\langle{x}\rangle_i]\,, \\
     \frac{\partial{\dot{x}_i}}{\partial{h_j}}\Bigr|_{\substack{i\neq j}} &= 0\,, \hspace{50pt}
     \frac{\partial{\dot{x}_i}}{\partial{h_i}} = -g(1-w)x_i\,, \\
     \frac{\partial{\dot{h}_i}}{\partial{x_j}}\Bigr|_{\substack{i\neq j}} &= 0\,, \hspace{50pt}
     \frac{\partial{\dot{x}_i}}{\partial{x_i}} = 1\,, \\
     \frac{\partial{\dot{h}_i}}{\partial{y_j}}\Bigr|_{\substack{i\neq j}} &= 0\,, \hspace{50pt}
     \frac{\partial{\dot{h}_i}}{\partial{y_i}} = -1\,.
\end{align*}
and evaluating them in a general $\fixp[{M}]{u}$ gives
\begin{equation}
    \mathbf{J} \equiv
    \left(
    \arraycolsep=5.4pt\def\arraystretch{1.4}
    \begin{array}{c|c}
    \bm{A} & \bm{B}  \\
    \hline
    \bm{I}_N  & -\bm{I}_N 
    \end{array}
    \right), 
    \qquad
    \textbf{A}=
    \left(
    \arraycolsep=5.4pt\def\arraystretch{2}
    \begin{array}{c|c}
    a_1 \bm{I}_{N-M} & \bm{0}_{N-M,M}  \\
    \hline
    a_2\bm{1}_{M,N-M}  & a_2(\bm{1}_M-\bm{I}_M) 
    \end{array}
    \right)
    \qquad
    \textbf{B}=
    \left(
    \arraycolsep=5.4pt\def\arraystretch{2}
    \begin{array}{c|c}
    \bm{0}_{N-M} & \bm{0}_{N-M,M}  \\
    \hline
    \bm{0}_{M,N-M}  & b\bm{I}_M 
    \end{array}
    \right),
\end{equation}
where 
\begin{equation}
\begin{aligned}
    a_1 &= \frac{1}{T_x} \cdot \left(1-\fixpval{x}[M] \cdot \dfrac{wM}{N-1}\right)\,,\qquad
    a_2 = -\frac{1}{T_x}\left(\fixpval{x}[M]\cdot\frac{w}{N-1}\right)\,,\\
    a_3 &= -(M-1)a_2\,,\qquad
    b = \frac{1}{T_x}(w-1) \cdot \fixpval{x}[M]\,.\\
\end{aligned}
\end{equation}
For the matrices $\mathbf{0}$ and $\mathbf{1}$ all entries are $0$ or $1$, respectively, while $\mathbf{I}$ denotes the identity matrix.

The structure of $\mathbf{A}$ and $\mathbf{B}$ block matrices depends on which particular fixed point from $\fixp[{M}]{u}$ is substituted into the Jacobian. Nevertheless, it suffices to make the calculations for only one conveniently ordered example to get the stability of all fixed points from that group.

To study stability, one must solve for the $2N$ number of eigenvalues:
\begin{equation} \label{eq:ap_charpoly1}
\det(\bm{J}-\lambda \mathbf{I}_{2N})=
\left(
\arraycolsep=5.4pt\def\arraystretch{1.4}
\begin{array}{c|c}
\mathbf{A} - \lambda \mathbf{I}_{N} & \mathbf{B}  \\
\hline
\mathbf{I}_N  & -\mathbf{I}_N - \lambda \mathbf{I}_{N}
\end{array}
\right) = 0\,.
\end{equation}
The determinant can be computed as products of the smaller submatrix determinants via Schur complements, therefore, eigenvalues as well as their algebraic multiplicities (degree of degeneracy) can be made clear if one factorizes the characteristic polynomial \label{eq:charpoly1} into the following terms:
\begin{equation} \label{eq:ap_charpoly2}
    \underbrace{(-1-\lambda)^{N-M}}_{\lambda_a} ~ \cdot ~ \underbrace{(a_1-\lambda)^{N-M}}_{\lambda_b} ~ \cdot ~ \underbrace{\biggl[(-1-\lambda) (b_\lambda - a_2) \biggr]^{M-1}}_{\lambda_c^{\pm}} ~ \cdot ~ \underbrace{(-1-\lambda)(b_{\lambda}-a_3)}_{\lambda_d^{\pm}} = 0\,,
\end{equation}
therefore
\begin{equation}  \label{eq:ap_eig}
\begin{aligned}
\lambda_i^{(a)} &= -1, ~ 1 \leq i \leq N-M, ~ \text{if } ~ M < N\,, \\
\lambda_i^{(b)} &= \frac{1}{T_x}(1 - w M \cdot \fixpval{x}[M]), ~ N-M+1 \leq i \leq 2(N-M), ~ \text{if } M < N\,,\\
\lambda_i^{(c)} &= \frac{1}{2} \left(-(1+a_2) \pm \sqrt{(1+a_2)^2 + 4(b - a_2)}\right), ~ 2(N-M)+1 \leq i \leq 2N-2, ~ \text{if} ~ M \geq 2\,,  \\
\lambda_i^{(d)} &= \frac{1}{2} \left(-(1+a_3) \pm \sqrt{(1+a_3)^2 + 4(b - a_3)}\right), ~ 2N-2 \leq i \leq 2N, ~ \text{if } M \neq 0\,. 
\end{aligned}
\end{equation}
Solving for each eigenvalue type yields
\begin{equation}  \label{eq:ap_eig}
\begin{aligned}
\lambda_a &= -1, \qquad m(\lambda_a) = N-M , \quad \text{if } ~ M < N\,,\\
\lambda_b &= \frac{1}{T_x}\left(1 - \frac{wM}{N-1} \cdot \fixpval{x}[M]\right), \qquad m(\lambda_b) = N-M, \quad \text{if } M < N\,, \\
\lambda_c^{\pm} &= \frac{1}{2} \left(-(1+a_2) \pm \sqrt{(1+a_2)^2 + 4(b - a_2)}\right), \qquad m(\lambda_c^{+}) = m(\lambda_c^{-}) = M-1, \quad \text{if} ~ M \geq 2\,, \\
\lambda_d^{\pm} &= \frac{1}{2} \left(-(1+a_3) \pm \sqrt{(1+a_3)^2 + 4(b - a_3)}\right), \qquad m(\lambda_d^{+}) = m(\lambda_d^{-}) = 1, \quad \text{if } M \neq 0\,,
\end{aligned}
\end{equation}
where with $m(\lambda)$ we denote the algebraic multiplicity of the eigenvalue with the condition that the respective eigenvalue type only exists if $m(\lambda)\geq 1$.

\begin{figure} [hbt!]
 \centering
 \includegraphics[width=\textwidth]{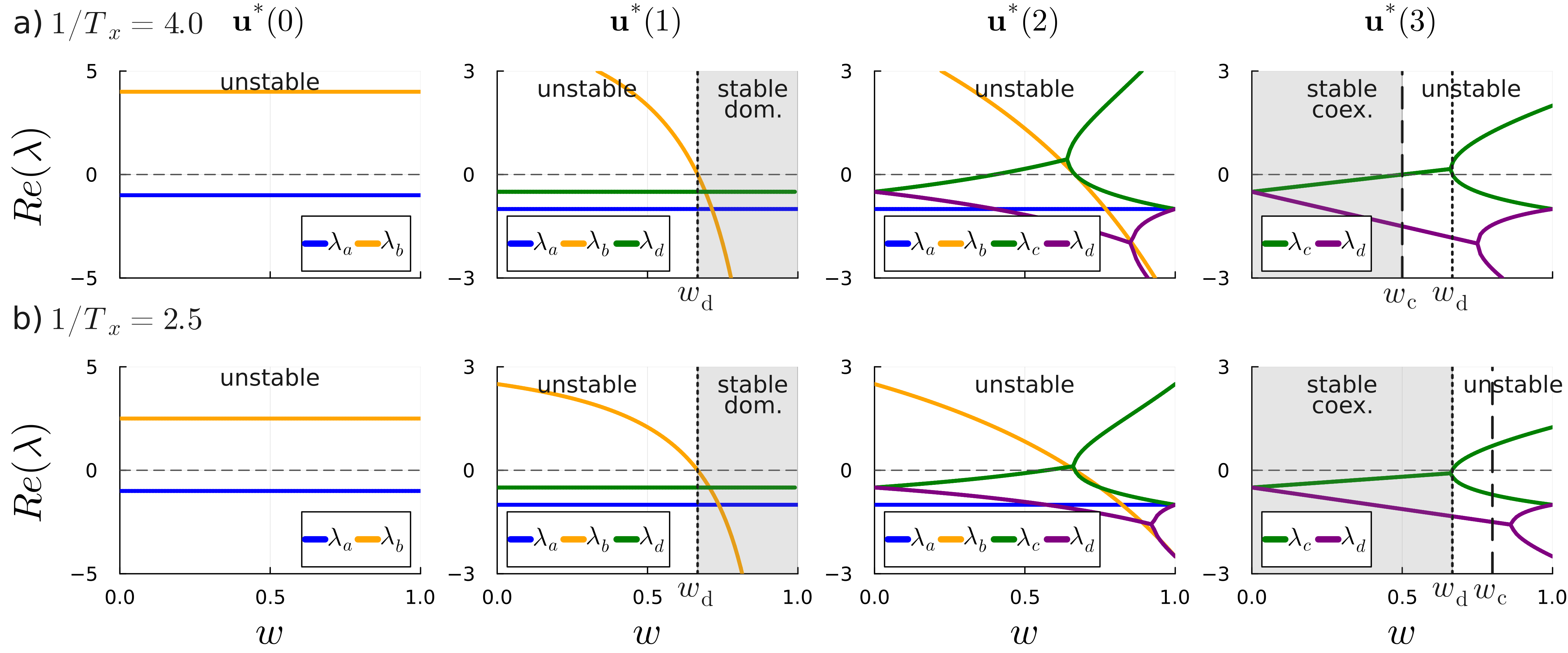}
 \caption{Real part of the eigenvalue spectrum, $\Re(\lambda)$, for $N=3$ items as a function of the competition strength $w$ for different fixed point types $\fixp[{M}]{u}$ with $M=0,1,2,3$ (columns from left to right). The gray-shaded region denotes the stable fixed point regimes (dominance and coexistence), delimited either by the heteroclinic bifurcation point at $w_\mathrm{d}$ (dotted vertical line) or by the Hopf-bifurcation point at parameter $w_\mathrm{c}$ (dashed vertical line). (a) In the top row for $1/T_x=4>N$, WLC emerges in the $w\in (w_\mathrm{c},w_\mathrm{d})$ interval. (b) In the bottom row for $1/T_x=2.5<N$, no WLC is possible as $w_\mathrm{d}<w_\mathrm{c}$.}
 \label{fig:eigenspectra} 
\end{figure}

In Fig.~(\ref{fig:eigenspectra}), an example of eigenvalue spectra is shown for the system with $N = 3$ competitors.

\paragraph{Chaotic dynamics.}
Our numerical analysis is based on the equivalence of
\begin{equation}
\label{eq:delay_split_orig} 
\dot{L}_i = r_p L_i \left( 1 - \frac{r_c}{K}Y_i - 
c \sum_{j\neq i}^N L_j \right), \qquad\quad
\dot{Y}_i = L_i - \alpha Y_i\,,
\end{equation}
and the delayed Lotka-Volterra system 
as defined by Eq.~(\ref{eq:rescaled})
(see also Sec.~3 in the main text).
In Eq.~(\ref{eq:delay_split_orig}), time is rescaled as $t_L = t / \alpha$, where $t$ denotes the time variable in the $(x,h)$ system~(\ref{eq:rescaled}), while $t_L$ denotes the time variable in the $(L,Y)$ formulation~(\ref{eq:delay_split_orig}).
The maximal Lyapunov exponents of the two corresponding systems are therefore related by 
\begin{equation}
\Lambda=\Lambda_L / \alpha\,,
\label{eq:lambdas}
\end{equation}
where the subscript $L$ denotes the Lyapunov exponent of the $(L,Y)$ system.

\begin{figure} [hbt!]
    \centering
    \includegraphics[width=0.8\linewidth]{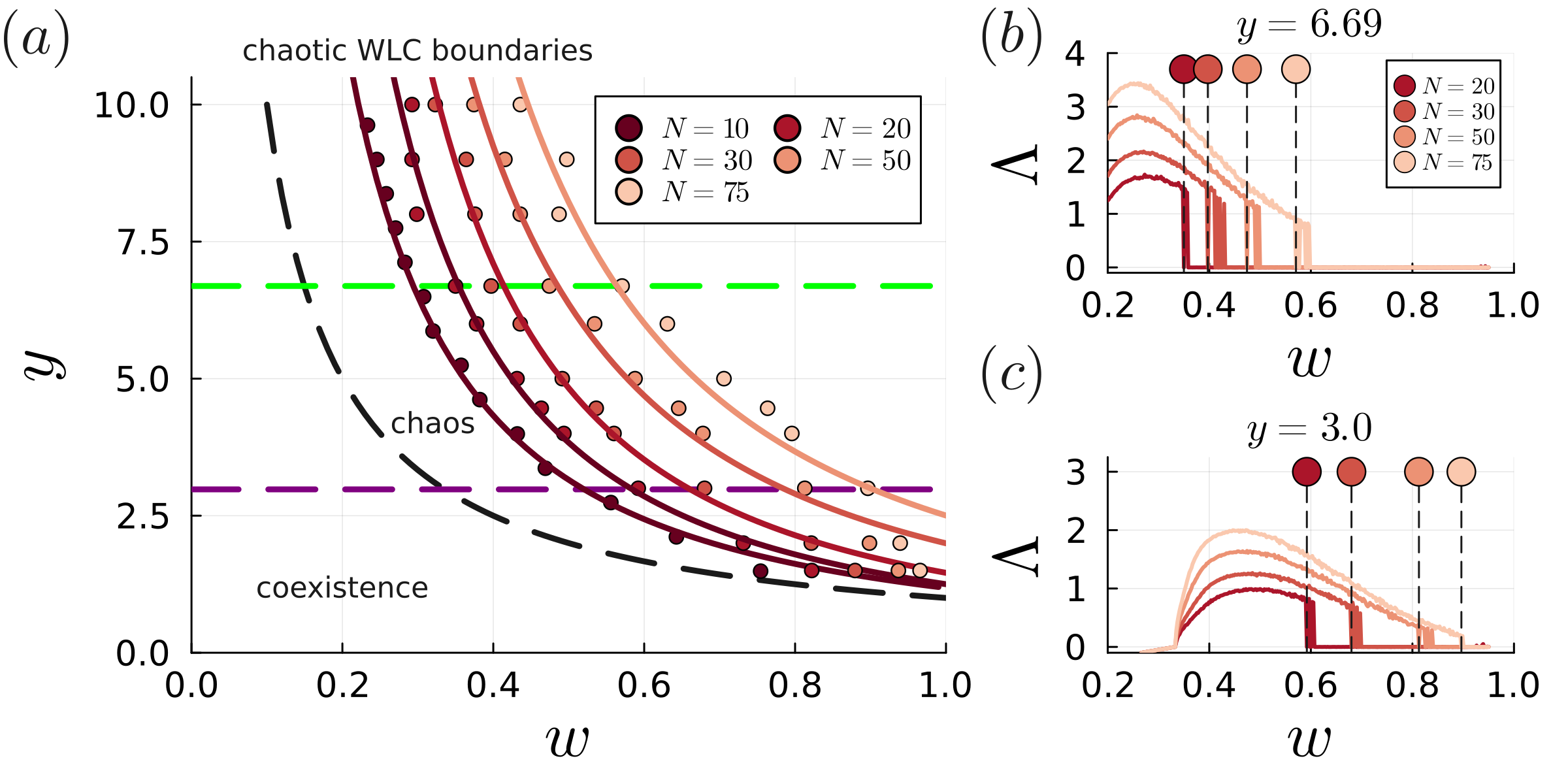}
    \caption{Extent of the parameter region where chaotic WLC occurs in systems of different sizes. (a) Schematic phase diagram showing the boundary of chaotic WLC regions for systems with $N \in \{10,20,30,50,75\}$ items. For every system size, the chaotic region is between the $1/w$ critical curve (black dashed curve) and the boundary curves shown with different shades of red, indicating that the chaotic region gets wider as $N$ increases. To obtain the boundary for each $N$, a few $y$ parameter levels were selected (points with circular markers), to which power law functions were fitted $y = 1/(T_x (N-1)) =  a \cdot w^{-\beta} $, shown by the solid curves, with $a \in \{1.18,1.25,1.46,1.99,2.50\}$ and $\beta \in \{-1.41,-1.61,-1.73,-1.67,-1.71\}$, respectively.  (b),(c) The largest Lyapunov exponent $\Lambda$ is shown in terms of $w$, for the two selected $y$ time scale separation levels, shown on panel (a) with green and purple dashed horizontal lines, respectively. The endpoints are marked when $\Lambda$ first falls below a threshold (near zero, indicating that after a transient, the system finally starts to converge to one of the limit cycles). For computing the maximal Lyapunov exponent, a transient time of $T_{L,{\mathrm{tr}}} = 5 \cdot 10^4 $ time units was used.}
    \label{fig:phase_diag_size_effect}
\end{figure}

The extent of the chaotic regime in the phase diagram of the reduced time scale $y=1/(T_x (N-1))$ and coupling strength $w$ is estimated through the computation of the largest Lyapunov exponent $\Lambda_L$ of system (\ref{eq:delay_split_orig}) for a large set of parameter pairs distributed on a grid covering the phase diagram. For intermediate system sizes, $N=20,30,50,75$, we computed $\Lambda$ via Eq.~(\ref{eq:lambdas}) for a dozen values of $1/(T_x (N-1))$ as a function of $w$, and defined the edge of the chaotic regime as the first value of $w$ for which $\Lambda=0$ (see Fig.~\ref{fig:phase_diag_size_effect}).

\newpage
%----------------------------------------
\section{Computing lifetime distributions for different system sizes and timescales}
\label{sec:ap_lifetimes}
%----------------------------------------

\subsection{Top-$k$ charts}

\begin{figure}[t]
     \centering
     \includegraphics[width=0.7\textwidth]{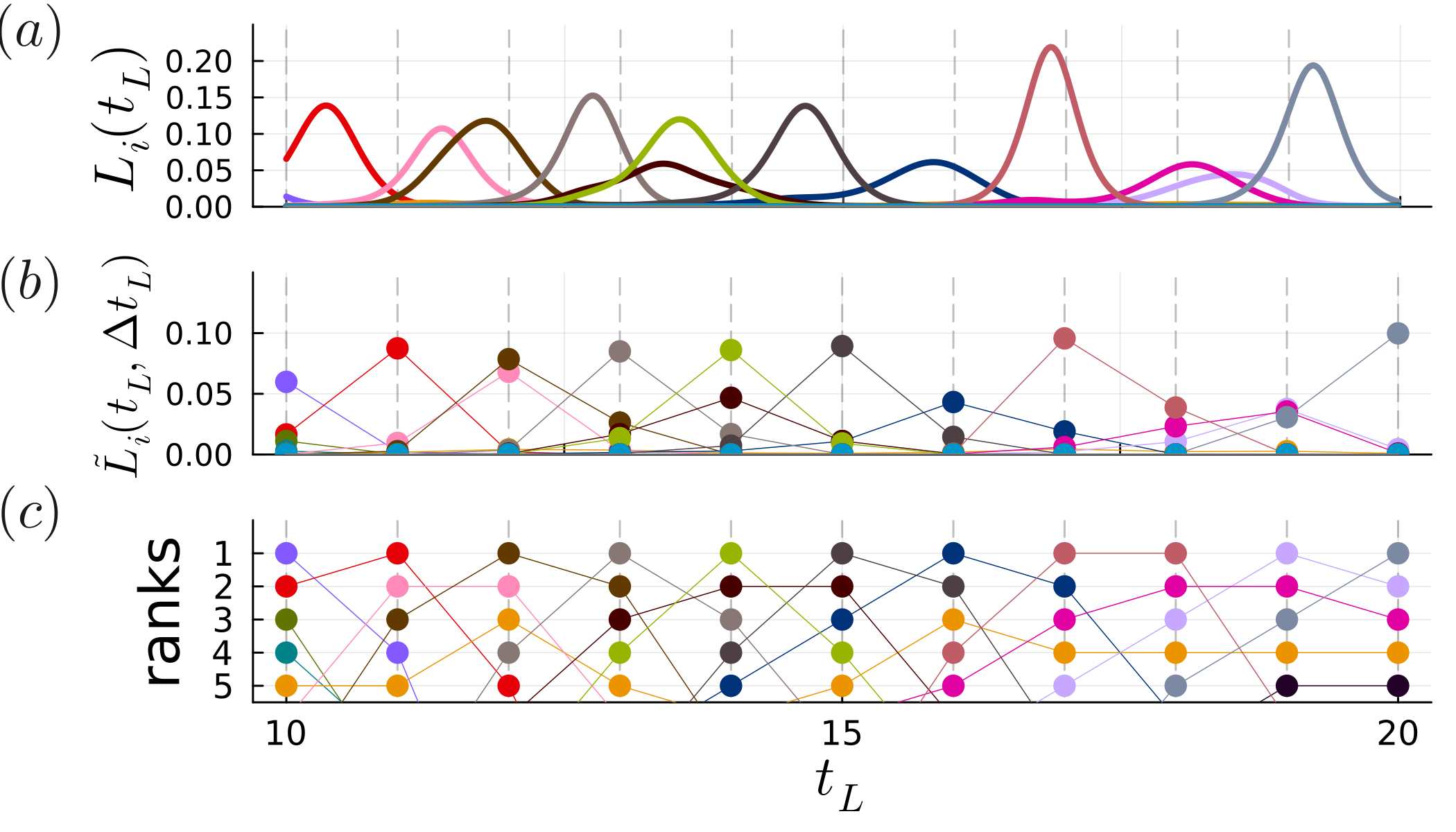}
\caption{\textbf{Constructing top-$k$ charts for competing
cultural items.} 
(a) A time series of $N=300$ topics $L_i$ in the 
winnerless chaotic regime for 
the parameters marked by B$_2$ in the phase
diagram of Fig.~(2) in the main text. The values
are $w\approx 0.264$ and $y \approx 6.69$, or, 
equivalently, $r = 10, K=1$, $c=2.4$ and $\alpha=0.005$,
as in \cite{attention_dynamics}.  
(b) Corresponding integrated popularity measures 
$\tilde{L}_i(t_L,\Delta t_L)$, see
Eq.~(18) and Eq.~(19), 
with $\Delta t_L=1$. 
(c) Top $k=5$ ranks in the popularity 
chart as a function of time $t_L$. The lifetime
$\tau = 1, 2, 3, ..$ of an item $i$ is given by the number 
of consecutive steps spent in the toplist, viz
when following the respective thin lines.} 
\label{fig:toplist}
\end{figure}

Assuming the activities $L_i(t)$ in Eq.~(\ref{eq:delay_split_orig}) serve as proxies for instantaneous attention (see Fig.~\ref{fig:toplist}(a))—representing metrics such as song plays or social media mentions as in \cite{attention_dynamics}— we define the overall popularity of item $i$ over an observation window $\Delta t_L$ from the recent past as
\begin{equation}
\tilde{L}_i(t_L,\Delta t_L)=\int_{t_L-\Delta t_L}^{t_L} L_i(\theta)~\mathrm{d}\theta\,,
\label{eq:popularity}
\end{equation}
where time $t_L$ is sampled now according to the window $\Delta t_L$ (see Fig.~\ref{fig:toplist}(b)).
In analogy with commercial cultural charts, we construct a "top-$k$" list at each interval, consisting of the $k$ items with the largest values of $\tilde{L}_i$ (for the time evolution of the top-5 ranks, see Fig.~\ref{fig:toplist}(c)). 

Here, we focus on the chart lifetime, $\tau$, defined as the number of consecutive windows of length $\Delta t_L$ during which an item remains in the top-$k$ chart. The actual time spent in the top-$k$ chart hence can be given as $\tau\cdot \Delta t_L$. The lifetime statistics can be characterized by the distribution $p(\tau)$, describing how often a particular lifetime $\tau$ is present during the attention dynamics. 

%%%%%%%%%%%%%%%%%%%%%%%%%%%%%%%%%%%%%%%%%

\subsection{System sizes}
\begin{figure} [hbt!]
    \centering    \includegraphics[width=0.9\linewidth]{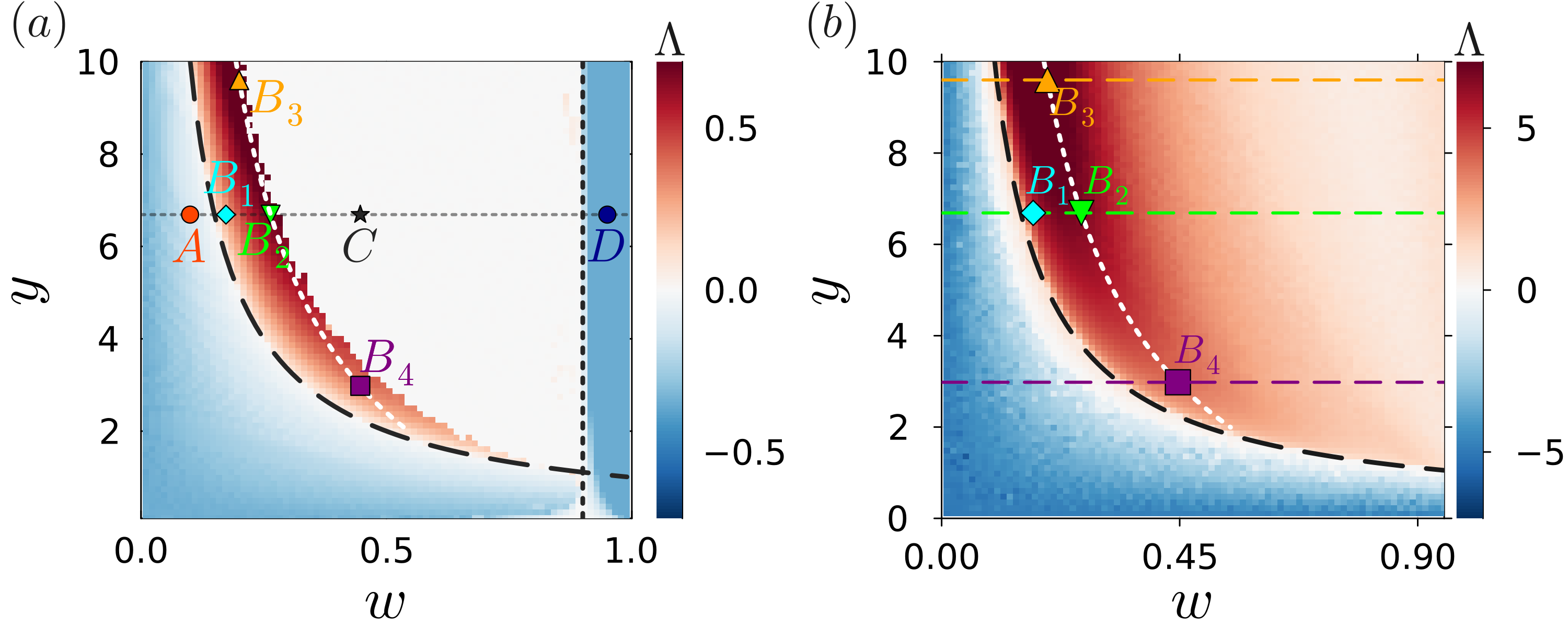}
    \caption{Phase diagrams as in Figs.~1 and 2 in the main paper, color-coded by the largest Lyapunov exponent, $\Lambda$, in the system-size–scaled parameter plane of time-scale separation and coupling strength, $(w,y)$, with $y=1/(T_x(N-1))$, for $N=10$ items in (a) and for $N=300$ items (b). The thick black dashed curve and the dotted vertical lines indicate the stability boundary of the coexistence and the WTA-type dominance fixed points, respectively (blue-shaded regions corresponding to stable regimes). 
    Specific parameter values used as examples for time series are indicated by the markers (and corresponding capital letters): in terms of the ($L,Y$) system parameters, $c$ is increasing along the horizontal dotted line, i.e.\ for the A, B$_1$, B$_2$, C, and D points, while $r=r_\mathrm{p}=r_\mathrm{c}$ is decreasing along the dotted curve inside the chaotic region, in the order of B$_3$, B$_2$, and B$_4$ points. All the other parameters, $N$, $K$, and $\alpha$ are kept constant along these curves.
    }
    \label{fig:phase_diagrams}
\end{figure}

Here, we complement the analysis presented in the main text by investigating the influence of additional model parameters on the lifetime distributions. We consider four different parameter sets in the rescaled phase diagram $(w,y)$ that correspond exactly to, or lie in the neighborhood of, those obtained by fitting various model-generated statistics to real data in~\cite{attention_dynamics} for $N=300$ topics (see the thin dashed curves in the phase diagrams shown in Fig.~\ref{fig:phase_diagrams}). Specifically:
\begin{itemize}
    \item first, along the $y = 1/(T_x(N-1))\approx 6.69$ horizontal dotted line in Fig.~\ref{fig:phase_diagrams}(a) the coupling strength $w$ is increased, which is equivalent in the $(L,Y)$ system (\ref{eq:delay_split_orig}) to increasing coupling $c$ with constant $r_p = r_c = r = 10$, with $c \in \{1.4, 2.4\}$ (corresponding to B$_1$ and B$_2$, respectively);
    \item second, along the $y = 1/(T_x(N-1))\propto 1/w - 1$ white dotted curve inside the chaotic region in Fig.~\ref{fig:phase_diagrams}(a) and (b), in terms of the $(L,Y)$ system (\ref{eq:delay_split_orig}), only $r_p = r_c = r$ is varied, with $r \in \{14.352, 4.456\}$, and constant $c = 2.4$ (corresponding to B$_3$ and B$_4$, respectively).
\end{itemize}
The parameter $K = 1.0$ is fixed, and it is identical in all cases. Furthermore, setting $\alpha = 0.166$ for $N = 10$ and $\alpha = 0.005$ for $N = 300$ yields identical values for the two system sizes in the reduced parameter space $(w, y)$, and hence correspond to the same points in the phase diagrams (compare panels (a) and (b) in Fig.~\ref{fig:phase_diagrams}).

\begin{figure} [t!]
    \centering
    \includegraphics[width=0.9\linewidth]{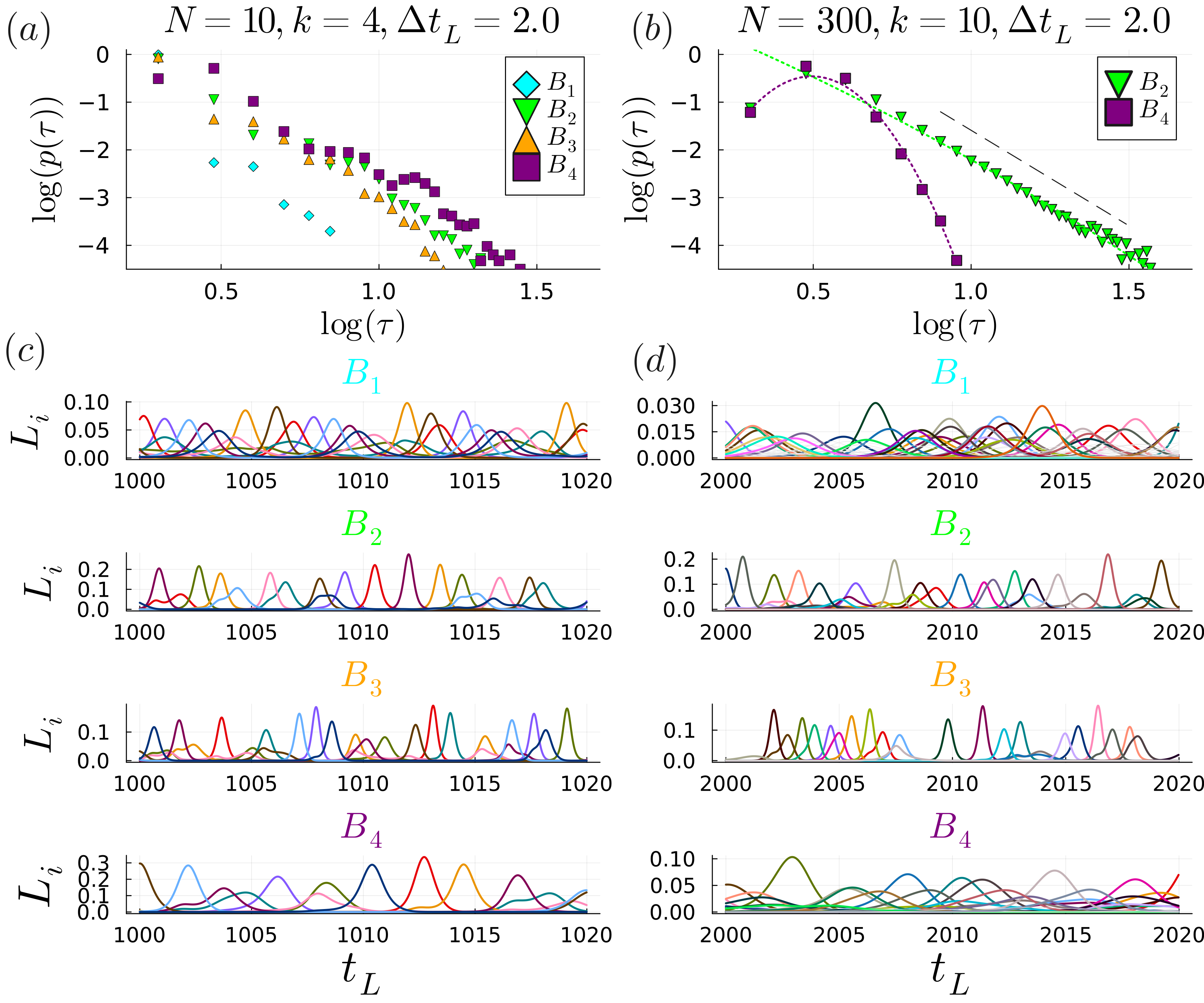}
    \caption{(a,b) Lifetime distributions $p(\tau)$, on a basis 10 log-log scale with $\Delta t_L=2$ for different parameter values in WLC regime (see the markers with the letters B$_1$, B$_2$, B$_3$, and B$_4$ in the phase diagrams shown in Fig.~\ref{fig:phase_diagrams}. \\
    (a) Distributions for $N=10$ items, built from charts of length $k=4$. Trajectories of $t_L = 6000$ time units were used with $20$ random initial conditions for each parameter, discarding the first $T_{L,\mathrm{tr}} = 1000$ time units (see examples of time series on panel~(c)). 
    (b) Distributions for $N=300$ items, built from charts of length $k=10$. The linear fit to the tail of the B$_2$ curve, corresponding to a power law distribution $\propto \tau^{-4}$ (black dashed line), and the parabolic fits $\sim q_0+(q_1-1)\log\tau+q_2(\log\tau)^2$ to both the B$_2$ and B$_4$ logarithmic data (dotted green and purple curves), indicating a log-normal distribution, are given as a guide to the eye.
    Trajectories of $t_L = 3000$ time units were used with $100$ random initial conditions for each parameter, discarding the first $T_{L,\mathrm{tr}} = 2000$ time units (for example, see panel~(d)). 
    (c,d) Time series of activities $L_i$ over 20 time units, corresponding to the four parameter sets above (from top to bottom). 
    (c) System with $N=10$ items. (d) System with $N=300$ items.
    Parameters: $K=1$ for all cases, $\alpha = 0.166$ for $N = 10$ and $\alpha = 0.005$ for $N = 300$, respectively. The coupling $c$ and production rate $r$ for the different points are given as follows: B$_1$: $r=10$, $c=1.4$; B$_2$: $r=10$, $c=2.4$; B$_3$: $r=14.532$, $c=2.4$; B$_4$: $r=4.456$, $c=2.4$.
    }
    \label{fig:lifetimes}
\end{figure}

\paragraph{$N=10$ topic system}
As a first example, we consider the WLC dynamics for the $N=10$ topic system for which the detailed stability and bifurcation analysis has been presented in the previous sections (see the corresponding phase diagram for $N=10$ on Fig.~\ref{fig:phase_diagrams}(a)). In the case of the periodic reactivation of popularities (parameter pair denoted by letter C on Fig.~\ref{fig:phase_diagrams}(a)), there is a single characteristic lifetime $\tau=4$, corresponding to $\tau\cdot\Delta t_L=8$ simulation time for remaining in the chart of the top $k=4$ topics (hence not shown here). 
On the other hand, in the chaotic region, the lifetimes are more widely distributed for all studied parameter values (see parameters B$_2$, B$_3$, and B$_4$ in Fig.~\ref{fig:lifetimes}(a)). The corresponding time series of activities show similar patterns of short activity bursts, with only a few topics dominating temporarily and occasionally longer-lasting items following the top scorer (see the flattened peaks for B$_2$ and B$_3$ in Fig.~\ref{fig:lifetimes}(c)). For a coupling strength $w$ close to the degenerate Hopf bifurcation point, short lifetimes dominate, leading to a rapidly decaying distribution (see the B$_1$ parameter in Fig.~\ref{fig:lifetimes}(a)). In this case, several topics are highly active at the same time (B$_1$ in Fig.~\ref{fig:lifetimes}(c)), leading to rapidly changing popularity rankings in the charts.

\paragraph{$N=300$ topic system}
As the number of topics increases, this picture becomes clearer (see the corresponding phase diagram for $N=300$ in Fig.~\ref{fig:phase_diagrams}(b)). As a second example, we consider the original parameters used in~\cite{attention_dynamics} with $N=300$ topics and charts of the $k=10$ most popular items. At the border of the stability curve in the phase diagram, the lifetime distribution reduces again to a delta-like peak (a single marker corresponding to B$_1$ in Fig.~\ref{fig:lifetimes}(b)), as only $\tau\le 2$ lifetimes are possible due to the rapidly changing rankings in the charts (for the corresponding time series, see B$_1$ in Fig.~\ref{fig:lifetimes}(d)). For parameters deeper in the chaotic region, with constant coupling strength $c=2.4$ (in terms of the $(L,Y)$ system parameters), we find two qualitatively distinct behaviors as the production rate $r$ is increased. For high production rates $r$ (resulting from the fitting of real data for recent years in \cite{attention_dynamics}), the lifetime distributions follow a relatively fast power law-like decay $\propto \tau^{-4}$ (see B$_2$ and B$_3$ in Fig.~\ref{fig:lifetimes}(b)), in agreement with our findings for recent Billboard and Spotify weekly charts in~\cite{Schneider2021}. On the other hand, for a lower cultural yield $r$, the distribution tends toward a log-normal form (B$_4$ in Fig.~\ref{fig:lifetimes}(b)), resembling a quadratic function in a log--log plot, similarly to the lifetime distributions of albums in the Billboard charts before the 1990s~\cite{Schneider2021}. \\

\subsection{Charting periods}
\begin{figure} [t!]
    \centering
    \includegraphics[width=0.9\linewidth]{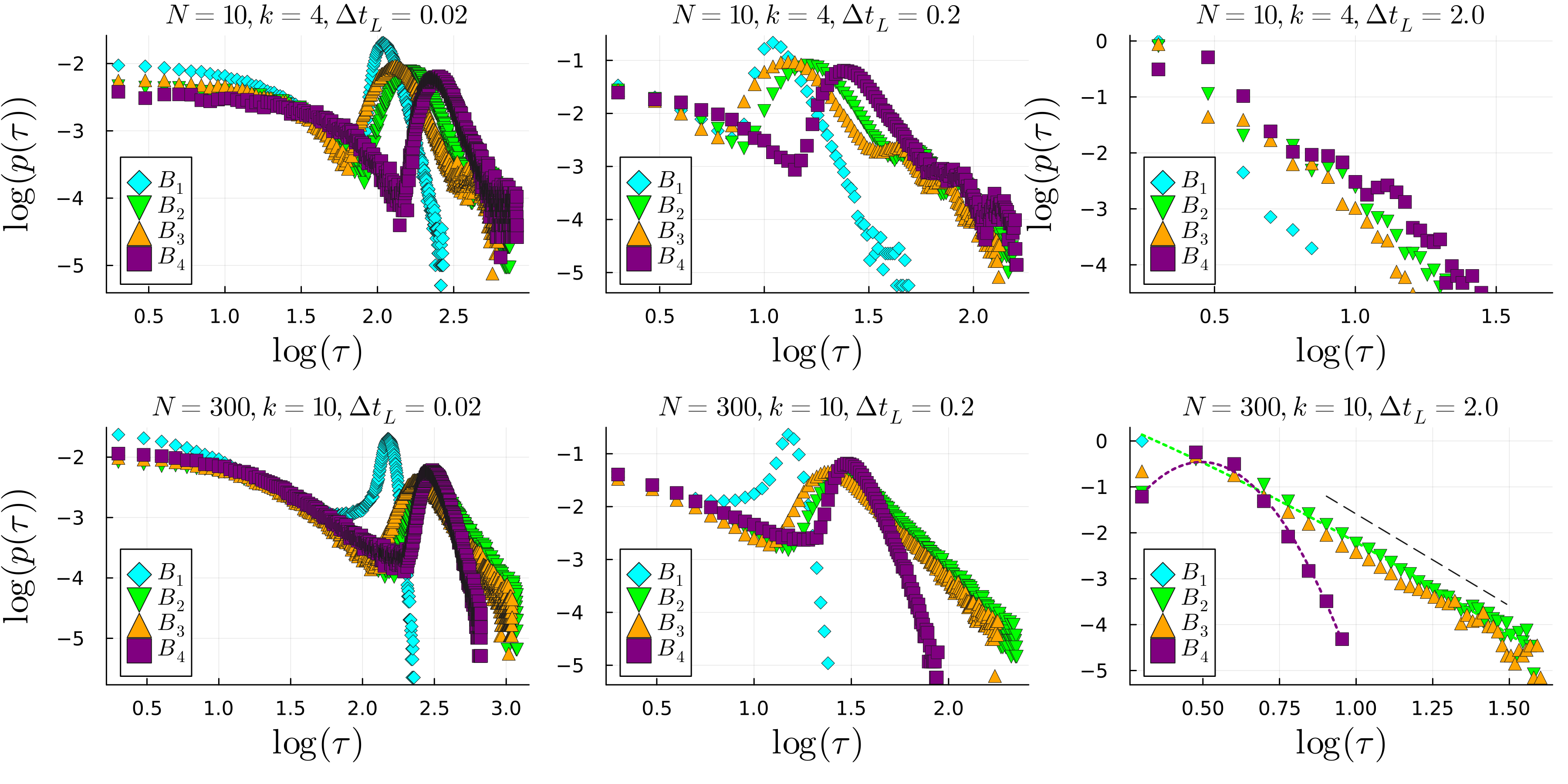}
    \caption{Lifetime distributions $p(\tau)$, on a basis 10 log-log scale for different parameter values in the WLC regime (see the markers with the letters B$_1$, B$_2$, B$_3$, and B$_4$ in the phase diagrams shown in Fig.~\ref{fig:phase_diagrams}). The time interval $\Delta t$ used to define the popularities increases from left to right (across columns), $\Delta t_L = 0.02, 0.2, 2.0$, respectively. \\
    \textit{Top row:} Distributions for $N=10$ items, built from charts of length $k=4$. Trajectories of $t_L=6000$ time units were used with $20$ random initial conditions for each parameter, discarding the first $T_{L,\mathrm{tr}}=1000$ time units.
    \textit{Bottom row:} Distributions for $N=300$ items, built from charts of length $k=10$. Trajectories of $t_L=3000$ time units were used with $100$ random initial conditions for each parameter, discarding the first $T_{L,\mathrm{tr}}=2000$ time units.  For the $\Delta t_L=2.0$ case, the linear fit to the tail of the B$_2$ curve, corresponding to a power law distribution $\sim \tau^{-4}$ (black dashed line), and the parabolic fits $\sim q_0+(q_1-1)\log\tau+q_2(\log\tau)^2$ to both the B$_2$ and B$_4$ logarithmic data (dotted green and purple curves), indicating a log-normal distribution, are given as a guide to the eye.
    }
    \label{fig:lifetimes_deltat}
\end{figure}

Toplists viewed on different timescales (daily, weekly, or monthly music album charts, or top hashtags, for example) can highlight different aspects of the underlying dynamics. When changing the charting time interval $\Delta t_L$, particularly when it is smaller than the average peak width, the lifetime distribution curves can be decomposed into three distinct parts (see Fig.~\ref{fig:lifetimes_deltat}).
\begin{enumerate}
    \item Short lifetimes reflect the microscopic characteristics of items overtaking one another, represented by the curve leading up to a local minimum (at $\tau \approx 2$ when $\Delta t_L = 0.02$, see the left column of Fig.~\ref{fig:lifetimes_deltat}), which is likely related to the specific shape of the $L_i$ fluctuations.
    \item A peak in the lifetime distribution, visible at small and intermediate timescales ($\Delta t_L = 0.02, 0.2$; see the left and middle columns of the same figure), which is associated with the average peak widths of $L_i$.
    \item A power-law or log-normal tail for long lifetimes, which is associated with single topics dominating the activity dynamics $L_i$. Note that this feature is present across all $\Delta t_L$ scales, but dominates when the charting interval is longer, and the peak of the distribution is shifted to the left (see the right column of the same figure).
\end{enumerate}

